\documentclass[11pt,a4paper]{article}
\pdfoutput=1
\usepackage{jcappub}
\usepackage{hyperref}
\usepackage{subfig}
\hypersetup{colorlinks = true, citecolor = blue}
\usepackage{cancel}

\begin{document}
%
%\preprint{ITP-UU-11/32, SPIN-11/24}
%
\title{
    Frame independent cosmological perturbations
       }
\author{Tomislav Prokopec}
\emailAdd{t.prokopec@uu.nl}
\author{and Jan Weenink}
\emailAdd{j.g.weenink@uu.nl}

\affiliation{Institute for Theoretical Physics and Spinoza Institute,\\
Utrecht University, Leuvenlaan 4, 3585 CE Utrecht, The Netherlands}
\abstract{
We compute the third order gauge invariant action
for scalar-graviton interactions in the Jordan frame.
We demonstrate that the gauge invariant action
for scalar and tensor perturbations on one physical hypersurface
only differs from that on another physical hypersurface
via terms proportional to the equation of motion
and boundary terms, such that the evolution of
non-Gaussianity may be called unique.
Moreover, we demonstrate that the
gauge invariant curvature perturbation and graviton
on uniform field hypersurfaces in the Jordan frame
are equal to their counterparts in the Einstein frame.
These frame independent perturbations are therefore
particularly useful in relating results in different
frames at the perturbative level. On the other hand,
the field perturbation and graviton
on uniform curvature hypersurfaces in the Jordan
and Einstein frame are non-linearly related, as are
their corresponding actions and $n$-point functions.
}
\arxivnumber{}
\maketitle

\section{Introduction}
The theory of General Relativity exhibits general covariance --
physical laws and observables are invariant under coordinate
reparametrizations. In a field-theoretic framework these
physical laws are derived from an action, written in
a manifestly covariant form in terms of tensors, covariant
derivatives, 4-vectors and scalars. Even though the action
is manifestly covariant, this is not at all clear
when the action is perturbed around a fixed background.
Here is where one runs into the gauge problem
of General Relativity -- the perturbations themselves are
dependent on the choice of coordinate system. As a consequence
it becomes complicated to extract physical observables from
the perturbed action.

Fortunately it is possible to specify unambiguous quantities
through the use of gauge invariant cosmological perturbations
\cite{Bardeen:1980kt}. This has been applied to the action
for a single scalar field in an expanding universe, by deriving
the manifestly gauge invariant quadratic action
for linearly gauge invariant scalar perturbations
\cite{Mukhanov:1981xt,Mukhanov:1990me}. In an inflationary
context the primordial power spectrum for these fluctuations
is predicted to be nearly scale invariant, which has recently been
established at the 5 $\sigma$ level by the Planck mission
\cite{Ade:2013rta}. Gauge invariant perturbations can also be
constructed at second order in perturbation theory by fixing
the gauge transformation order by order
\cite{Acquaviva:2002ud,Bartolo:2004if,Malik:2003mv,Malik:2008im}.
Moreover, in Ref. \cite{Maldacena:2002vr} the third order
action for standard single field inflation was derived in different
gauges, while Refs. \cite{Seery:2005wm,Chen:2006nt} derived the action
for more general single field inflation in the uniform field gauge.
Finally the fourth order action was computed in the uniform
curvature gauge in \cite{Sloth:2006az,Seery:2006vu}.
Only recently the third order action for scalar perturbations was
treated in a completely gauge invariant way \cite{Rigopoulos:2011eq,Prokopec:2012ug},
and it was demonstrated how to find the gauge invariant vertices
and second order gauge invariant variables from the action.

Closely related to the coordinate invariance of General Relativity
and the associated difficulties at the perturbative level,
is the case of frame transformations and physical equivalence.
A single scalar field may be non-minimally coupled to the Ricci scalar in the action,
which is then said to be in the \textit{Jordan frame}. It has been long known
that such a theory may be rewritten into the
standard, minimally coupled \textit{Einstein frame} form by
redefining the metric and scalar field. Since these are just field redefinitions,
no physical content is lost in the frame transformation, and
the Jordan and Einstein frame are in that sense physically equivalent.
This proves very convenient in finding Jordan frame results \textit{via}
the simpler Einstein frame. For example, the concept of Higgs inflation
\cite{Salopek:1988qh,Bezrukov:2007ep} (see also \cite{Futamase:1987ua,Fakir:1990eg}),
where the Higgs field is coupled
to the Ricci scalar by a large non-minimal coupling, is most easily
illustrated in the Einstein frame, where the potential becomes exponentially
flat for large field values.

Although the equivalence between the Jordan and Einstein
frame is straightforwardly established for the complete action,
this is not obvious for the perturbed action. The reason is that
the perturbations in one frame do not coincide with those
in the other frame, because the frame transformations are
field dependent (note the similarity to the gauge dependent case).
As such it becomes troublesome to relate certain
quantity in one frame, for example an $n$-point function in the Jordan
frame, to the corresponding quantity in the other frame.

Luckily, also here it is possible to define unambiguous perturbations
which take the same form in either frame. We call these perturbations
\textit{frame independent cosmological perturbations}.
In Refs. \cite{Makino:1991sg,Fakir:1992cg} the frame independent
scalar perturbation was derived, and it was shown that it coincides
with the linear gauge invariant curvature perturbation. In terms
of such a frame independent perturbation the equivalence between
Jordan and Einstein frame is almost trivially established for the
second order action, and it was used to derive the power spectrum
in the Jordan frame from the well-known Einstein frame result.
Also, the second order action was derived directly in the Jordan
frame, and it was shown to be related to the Einstein frame action
via a frame transformation \cite{Weenink:2010rr}.
Furthermore, the frame independence of the gauge invariant curvature
perturbation has been demonstrated to second order, and was used
to compute $f_{\rm NL}$ in the Jordan frame via the Einstein frame
\cite{Sugiyama:2010zz}. The action for third order scalar perturbations
in the Jordan frame was computed in Ref. \cite{Qiu:2010dk}
in the uniform field gauge, and it was shown
in Ref. \cite{Kubota:2011re} that it can be derived from the Einstein
frame action by frame transformations of the background quantities.
In recent work \cite{Prokopec:2012ug} we took a gauge invariant
approach and showed how to derive the cubic Jordan frame action
for gauge invariant scalar perturbations from the Einstein frame.

In most of the works above the only perturbation considered
was the scalar degree of freedom in the action.
In this work we extend our previous work to include the graviton
and scalar-graviton interactions in the action. We perform
all computations directly in the Jordan frame with a general
non-minimal coupling. In section \ref{sec: The model} we
introduce the model and perturbations. In section
\ref{sec: The gauge invariant action in the Jordan frame}
we compute the gauge invariant cubic vertices for
the graviton and for scalar-graviton interactions
on different hypersurfaces. Finally, in section
\ref{sec: Frame independent cosmological perturbations}
we show how the gauge invariant cubic action in the
Jordan frame is related to that in the Einstein frame.

\section{The model}
\label{sec: The model}
We consider a single scalar field in the Jordan frame,
\textit{i.e.} coupled non-minimally to the Ricci scalar,
\begin{equation}
S=\frac12\int d^4 x \sqrt{-g}\left\{R F(\Phi)
-g^{\mu\nu}\partial_{\mu}\Phi\partial_{\nu}\Phi -2V(\Phi)\right\}
\,.
\label{nonminimalaction}
\end{equation}
The function $F(\Phi)$ is a general function of the scalar
field $\Phi$. For a minimally coupled theory $F(\Phi)=M_{P}^2\equiv 1$,
where $M_P$ is the reduced Planck mass. In the case of
Higgs inflation the coupling is $F(\Phi)=M_P^2-\xi \Phi^2$.

We are interested in finding the action \eqref{nonminimalaction}
up to third order in perturbations. Since the theory of general
relativity is covariant, some of the degrees of freedom in the
action are not physical. In order to eliminate unphysical degrees
of freedom from the action, and to separate dynamical from
constraint degrees of freedom, it is most convenient to slice
up spacetime into spatial hypersurfaces. This can be done
using the Arnowitt-Deser-Misner (ADM) metric \cite{Arnowitt:1962hi},
which is defined through the line element
\begin{equation}
ds^2=-N^2dt^2+g_{ij}(dx^i+N^idt)(dx^j+N^jdt)
\,.
\label{ADMlineelement}
\end{equation}
Here $g_{ij}$ is the spatial metric and $N$ and $N^{i}$
are the lapse and shift functions, respectively.
In terms of the ADM metric the action \eqref{nonminimalaction}
can be written (up to boundary terms) as
\cite{Weenink:2010rr}
\begin{align}
S=\frac12 \int d^3xdt \sqrt{g}\Biggl\{&
N R F(\Phi)+\frac{1}{N}\left(E^{ij}E_{ij}-E^2\right)F(\Phi)
-\frac{2}{N} E F'(\Phi)\left(\partial_t{\Phi}
- N^{i}\partial_i\Phi\right)
\nonumber\\
&+2g^{ij}\nabla_iN \nabla_jF(\Phi)
+\frac{1}{N}\left(\partial_t{\Phi}-N^{i}\partial_i\Phi\right)^2
-Ng^{ij}\partial_i\Phi\partial_j\Phi-2 N V(\Phi)\Biggr\}
\,,
\label{ADMactionnonminimalEij}
\end{align}
where $F'(\Phi)=dF(\Phi)/d\Phi$ and
the measure $\sqrt{g}$, the Ricci scalar $R$
and covariant derivatives $\nabla_i$ are composed
of the spatial part of the metric $g_{ij}$ alone.
The quantities $E_{ij}$ and $E$ are related
to the extrinsic curvature $K_{ij}$ as $E_{ij}=-NK_{ij}$,
with
\begin{align}
E_{ij}&=\frac{1}{2}\left(\partial_tg_{ij}-\nabla_iN_j-\nabla_jN_i\right)\\
E&=g^{ij}E_{ij}
\,.
\end{align}
As one observes from Eq. \eqref{ADMactionnonminimalEij},
there are no kinetic-like terms for $N$ and $N^{i}$.
These fields therefore act as constraint or auxiliary fields.
The remaining dynamical variables are $g_{ij}$ and $\Phi$.
Although they contain in principle seven dynamical degrees of freedom,
only three of them are physical due to the additional gauge
freedom in the action. Diffeomorphism invariance (or covariance)
allows one to reparametrize coordinates, which means
that four degrees of freedom are unphysical. They can be
eliminated from the action by (a): fixing the gauge freedom,
or (b): constructing gauge invariant perturbations.
Gauge fixing has the advantage that a clever gauge choice
can lead to results much faster.
A problem is that there can be gauge artifacts due to
incomplete gauge fixing. Also, it may prove difficult to relate
results in one gauge to those in another, as is the case
in gravity where gauge transformations are non-linear.
The gauge invariant approach is on the one hand more tedious,
as one should take all degrees of freedom into account
and rewrite \textit{e.g.} the action or equations of motion
in terms of gauge invariant variables. On the other hand,
it is more thorough and one is guaranteed to obtain
results for the physical variables.

In this work we take a mixed approach. We fix the
gauge freedom with regards to spatial coordinate transformations, but
treat time reparametrizations in a gauge invariant way.
As is well known from work on first order scalar perturbation theory
(and the second order action describing these perturbations)
this describes dynamical scalar and tensor perturbations that are gauge invariant,
and constraint perturbations that are partially gauge fixed.
Thus, this approach treats the most relevant perturbations,
that is the dynamical perturbations, in a gauge invariant manner.

Our goal is to construct the third order action for
gauge invariant scalar and tensor perturbations, starting
from the unperturbed Jordan frame action \eqref{ADMactionnonminimalEij}.
The perturbations are considered on top of the homogeneous
FLRW background.
\begin{align}
g_{ij}&=a^2e^{2\zeta}\left(e^{\alpha}\right)_{ij}
\nonumber\\
\Phi&=\phi+\varphi
\nonumber\\
N&=\bar{N}\left(1+n\right)
\nonumber\\
N^i&=a^{-1}\bar{N}(t)(a^{-1}\partial_i s+n_{i}^{T})
\label{perturbations}
\,,
\end{align}
where $\partial_in_{i}^{T}=0$
and the background quantities are $a=a(t)$, $\phi=\phi(t)$ and $\bar{N}=\bar{N}(t)$
\footnote{Some notes on conventions, such as indices placement and factors of $\bar{N}$
and $a$, as well as some intermediate expansions of quantities in \eqref{ADMactionnonminimalEij}
can be found in appendix \ref{sec: Conventions and expansions}.}.
This background can be inserted in the action \eqref{ADMactionnonminimalEij}.
The background Friedmann equations and field equations are obtained by
a variation of this action with respect to $\bar{N}$, $a$ and $\phi$.
This gives, respectively,
\begin{align}
H^2&=\frac{1}{6F}\left[\dot{\phi}^2+2V-6H\dot{F}\right]
\nonumber\\
\dot{H}&=\frac{1}{2F}\left(-\dot{\phi}^2+H\dot{F}-\ddot{F}\right)
\nonumber\\
0&=\ddot{\phi}+3H\dot{\phi}+V'-6\left(2 H^2+\dot{H}\right)\frac12F'
\label{fieldequation}
\,.
\end{align}
Here we have defined the dotted derivative as $\dot{a}\equiv da/(\bar{N}dt)$,
and the Hubble parameter is $H=\dot{a}/a$.

Next we consider perturbations. As we are interested in finding
the third order action, all perturbations are of second order.
Any third order perturbation vanishes due to the background
equations of motion. The perturbation $\alpha_{ij}$ is part of the
scalar-vector-tensor decomposition of the metric \eqref{perturbations}
and reads
\begin{align}
\alpha_{ij}&=\frac{\partial_i\partial_j\tilde{h}}{a^2}
+\frac{\partial_{\left(i\right.}h_{\left.j\right)}^T}{a}
+\gamma_{ij}
\,,
\label{tensormetricperturbation}
\end{align}
with
\begin{align}
\partial_i h_i^T&=0
\,,
\qquad
\partial_i\gamma_{ij}=0
\,,
\qquad
\gamma_{ii}=\delta_{ij}\gamma_{ij}=0
\,.
\end{align}
The perturbation $\gamma_{ij}$ is thus the transverse
traceless tensorial perturbation, and is often
referred to as the graviton.
Even though the complete action \eqref{nonminimalaction}
is generally covariant, the perturbations themselves are
not. An infinitesimal diffeomorphism generated
by a vector field $\xi^{\mu}=(\xi^0,\xi^{i})$,
$\xi^{i}=a^{-1}\partial_i\tilde{\xi} +\xi_{i}^{T}$, induces a gauge
transformation of the perturbed metric and scalar field,
which for the particular (dynamical) perturbations in Eqs.
\eqref{perturbations} and \eqref{tensormetricperturbation}
means that to first order
\begin{align}
\varphi & \rightarrow~\varphi+\dot{\phi}\frac{\xi^0}{\bar{N}}
\nonumber\\
\zeta & \rightarrow~\zeta+H\frac{\xi^0}{\bar{N}}
\nonumber\\
\frac{\partial_i\partial_j\tilde{h}}{a^2} & \rightarrow
~\frac{\partial_i\partial_j\tilde{h}}{a^2}
-\frac{2\partial_i\partial_j\tilde{\xi}}{a^2}
\nonumber\\
\frac{\partial_{\left(i\right.}h^T_{\left.j\right)}}{a} & \rightarrow
~\frac{\partial_{\left(i\right.}h^T_{\left.j\right)}}{a}
-\frac{2\partial_{\left(i\right.}\xi^T_{\left.j\right)}}{a}
\nonumber\\
\gamma_{ij} & \rightarrow ~\gamma_{ij}
\,.
\label{firstordergaugetransformations}
\end{align}
Here is where the partial gauge fixing comes in.
The spatial gauge freedom can be fixed by
setting
\begin{equation}
\tilde{h}=0\,,
\qquad\qquad
h_i^T=0
\,,
\label{spatial gauge fix}
\end{equation}
such that the perturbed metric
can be written as $g_{ij}=a^2e^{2\zeta}(e^{\gamma})_{ij}$
\footnote{The gauge fixing essentially
means that we choose $\tilde{\xi}=\frac12\tilde{h}+\mathcal{O}(\xi^2)$ and
$\xi_i^T=\frac12 h_i^T+\mathcal{O}(\xi^2)$. Therefore, in any other place
where $\xi_i^T$ and $\tilde{\xi}$ appear, such as the
second order gauge transformations of $\zeta$, $\varphi$
and $\gamma_{ij}$, or first order transformation of
the constraint fields, the perturbations $\tilde{h}$
and $h_i^T$ reappear. However, in this work we
actually set $\tilde{h}$ and $h_i^T$ to zero everywhere.
In a sense, we thus only consider temporal gauge
transformations.}.
The only remaining gauge freedom is the temporal gauge
freedom, characterized by the gauge parameter $\xi^0$.
At first order in perturbations the graviton $\gamma_{ij}$
is gauge invariant, but the scalar perturbations $\zeta$
and $\varphi$ are not. However, it is possible to construct
a scalar combination which is gauge invariant to first order.
At second order both the graviton and scalar perturbations
are gauge non-invariant, but also here it is possible
to define non-linear variables which are gauge invariant
to second order. In principle there are infinitely many
gauge invariant variables at second order, but they are all
related by non-linear transformations.
In a previous work \cite{Prokopec:2012ug} we focused on the
gauge invariant scalar perturbations and showed how
to construct the gauge invariant action at third order.
In this work we shall focus mostly on the graviton and
scalar-graviton interactions.

\section{The gauge invariant action in the Jordan frame}
\label{sec: The gauge invariant action in the Jordan frame}

We now wish to find the gauge invariant action at third order
for the dynamical scalar and tensor. Of course, the action
\eqref{ADMactionnonminimalEij} is gauge invariant by definition,
since it originates from the manifestly covariant action \eqref{nonminimalaction}.
However, the perturbations \eqref{perturbations}
are not gauge invariant \eqref{firstordergaugetransformations}. Thus,
we set out to find the \textit{manifestly gauge invariant action at the
perturbative level}.

In order to do so, we first have to deal with the lapse function
$N$ and shift vector $N^{i}$. As was mentioned in the previous section,
these fields act as auxiliary fields. Thus, they can be solved for
and their solution can be inserted back into the action\cite{Maldacena:2002vr}.
For the second order action, it is only necessary to find the solution
for $N$ and $N^{i}$ to first order in perturbations. Any second
order term for $N$ multiplies the background equation for $H^2$
\eqref{fieldequation}, as this
equation is precisely derived by varying the action with respect
to $N$. A second order perturbation of $N^{i}$ appears as a total
spatial derivative, and such a term vanishes likewise.

For the third order action, it is also sufficient to know
the solutions for $N$ and $N^{i}$ to first order in perturbations.
The third order terms vanish due to the background equations
of motion, and the second order perturbations of $N$ and $N^{i}$
multiply the first order solutions of the constraint equations.

The first order solutions for $N$ and $N^{i}$ are found from
the equations of motion for these variables. Starting from
the action \eqref{ADMactionnonminimalEij} we find
\begin{align}
0=&\frac12 \sqrt{g}\biggl\{
[R-N^{-2}(E^{ij}E_{ij}-E^2)]F(\Phi)+2N^{-2}EF'(\Phi)(\partial_t\Phi-N^{i}\partial_i\Phi)
-2g^{ij}\nabla_i\nabla_jF(\Phi)
\nonumber\\
&\qquad~-N^{-2}(\partial_t\Phi-N^{i}\partial_i\Phi)^2
-g^{ij}\partial_i\Phi\partial_j\Phi-2 V(\Phi)\biggr\}
\nonumber\\
0=&\nabla_j\left[N^{-1}(E^{jk}g_{ik}-E\delta_{ij})F(\Phi)
-\delta_{ij}N^{-1}F'(\Phi)(\partial_t\Phi-N^{i}\partial_i\Phi)\right]
\nonumber\\
&-N^{-1}(-EF'(\Phi)+\partial_t\Phi-N^{i}\partial_i\Phi)\partial_i\Phi
\,.
\label{constraintequationsNi:nonminimal}
\end{align}
The second equation gives, to first order in perturbations
\begin{align}
0=&\partial_i
\left[2(2HF+\dot{F})n-4F\dot{\zeta}-2\dot{\phi}\varphi
-2F'\dot{\varphi}+2HF'\varphi-2F''\dot{\phi}\varphi\right]
-\frac12 \frac{\nabla^2}{a}n^{iT}
\,,
\end{align}
which present the solutions
\begin{align}
n=& \frac{1}{2HF+\dot{F}}\left[2F\dot{\zeta}+\dot{\phi}\varphi
+F'\dot{\varphi}-HF'\varphi+F''\dot{\phi}\varphi\right]
\nonumber\\
n_i^{T}=&0
\,.
\label{lapsesolution:nonminimal}
\end{align}
The other constraint in Eq. \eqref{constraintequationsNi:nonminimal}
can also be expanded to first order in perturbations, and this
presents the solution for the scalar shift perturbation $\nabla^2 s$.
By using the equations of motion \eqref{fieldequation}
and the first order solution for $n$ we find
\begin{align}
\frac{\nabla^2s}{a^2}=&
\frac{-1}{H+\frac12\frac{\dot{F}}{F}}
\frac{\nabla^2}{a^2}
\left(\zeta+\frac12\frac{F'}{F}\varphi\right)
+\frac{1}{2F}\frac{\dot{\phi}^2+\frac{3}{2}\frac{\dot{F}^2}{F}}
{\left(H+\frac12\frac{\dot{F}}{F}\right)^2}
\left(\zeta-\frac{H}{\dot{\phi}}\varphi\right)^{\cdot}
\nonumber\\
=& \frac{\nabla^2}{a^2}\left[\frac{-1}{H+\frac12\frac{\dot{F}}{F}}
\left(\zeta+\frac12\frac{F'}{F}\varphi\right)+\chi\right]
\,.
\label{shiftsolution:nonminimal:b}
\end{align}
Here we have implicitly defined the scalar quantity $\chi$.
The solutions \eqref{lapsesolution:nonminimal} and \eqref{shiftsolution:nonminimal:b}
may be inserted into the action \eqref{ADMactionnonminimalEij}
perturbed to second and third order. The remaining terms in the action only depend on
$\varphi$, $\zeta$ and $\gamma_{ij}$
\footnote{It is possible to keep the other scalar and vectorial part
of the metric perturbation, $\tilde{h}$ and $h_i^T$, see Eq. \eqref{tensormetricperturbation}
In that case the first order solutions of the transverse part of
the shift is $n_i^T/a=\frac12(h_i^T/a)^{\cdot}$, and the longitudinal
part gets an extra contribution $\nabla^2 s/a^2=\frac12 (\nabla^2 \tilde{h}/a^2)^{\cdot}+\ldots$.
When these solutions are inserted in the second order perturbed action,
the quadratic parts in $\tilde{h}$ and $h_i^T$ disappear completely.
This shows that these variables are not propagating degrees of freedom.
}.

Before we start to compute the gauge invariant action for
the dynamical perturbations, let us note that there are
alternative ways to deal with the non-dynamical degrees
of freedom $N$ and $N^{i}$, see Refs. \cite{Prokopec:2010be}
and \cite{Weenink:2010rr}. At second order the perturbations
of $N$ and $N^{i}$ can be decoupled from the rest of the
action (essentially "completing the square"). As a consequence
extra terms appear in the second order action for the remaining
dynamical variables, which, not surprisingly, are precisely
those that one would get after replacing $n$ and $s$ by their
first order solutions. The decoupled perturbations are gauge
invariant and their equation of motion is the first order
solution of the constraint equation. At third order
the perturbations for $N$ and $N^{i}$ can also be decoupled,
with the decoupled fields giving the second order solutions
for the constraint equations, and extra terms in the third order
action corresponding to the substitution of the first order
solutions of the constraint fields in the action. The procedure
is explained in appendix \ref{sec: Decoupling the constraint fields in the action}.

\subsection{Second order gauge invariant action}
\label{sec: Second order gauge invariant action}
Here we briefly review the quadratic computation, which was
done before in the Jordan frame \cite{Weenink:2010rr}.
The scalar field action in the Jordan frame \eqref{ADMactionnonminimalEij}
can be expanded to second order by inserting the perturbations
\eqref{perturbations}. Details of expansions of individual terms
can be found in appendix \ref{sec: Conventions and expansions}.
Next, the perturbations
of $n$ and $s$ can be replaced by its first order solutions
\eqref{lapsesolution:nonminimal} and \eqref{shiftsolution:nonminimal:b}.
This immediately yields the gauge invariant action for
the graviton,
\begin{equation}
S^{(2)}[\gamma^2]=\frac12\int d^3xdt \bar{N} a^3\frac{F}{4}
\left[\dot{\gamma}_{ij}\dot{\gamma}_{ij}
-\left(\frac{\partial \gamma_{ij}}{a}\right)^2\right]
\,.
\label{nmaction:2:gammagamma}
\end{equation}
Any cross terms $\varphi\gamma$ and $\zeta\gamma$ vanish as total derivatives.
The remaining terms in the action are quadratic $\zeta^2$ and $\varphi^2$,
or mixed terms $\zeta\varphi$. They may be combined into
the action for a single scalar perturbation,
\begin{align}
S^{(2)}[w_{\zeta}^2]=\frac12\int d^{3}xdt \bar{N} a^{3}
\frac{\dot{\phi}^2+\frac{3}{2}\frac{\dot{F}^2}{F}}
{\left(H+\frac12\frac{\dot{F}}{F}\right)^2}
\left[\dot{w}_{\zeta}^2-\left(\frac{\partial w_{\zeta}}{a}\right)^2\right]
\,,
\label{nmaction:2:wzetawzeta:GI}
\end{align}
where $w_{\zeta}$ is defined as
\begin{equation}
w_{\zeta}=\zeta-\frac{H}{\dot{\phi}}\varphi
\,.
\label{wzeta:firstorder}
\end{equation}
This variable is gauge invariant to first order in perturbations,
as can be checked straightforwardly from Eq. \eqref{firstordergaugetransformations}.
$w_{\zeta}$ is commonly called the \textit{curvature perturbation
on uniform field hypersurfaces}, since it reduces to $\zeta$ in the gauge $\varphi=0$.
Alternatively, we can define another gauge invariant variable,
the \textit{field perturbation on uniform curvature hypersurfaces}, by
\begin{equation}
w_{\varphi}=\varphi-\frac{\dot{\phi}}{H}\zeta
\,.
\label{wvarphi:firstorder}
\end{equation}
Also this variable is gauge invariant, and in terms of it
the action can be written as
\begin{align}
S^{(2)}[w_{\varphi}^2]=\frac12\int d^{3}xdt \bar{N} a^{3}
\frac{\dot{\phi}^2+\frac{3}{2}\frac{\dot{F}^2}{F}}
{\left(H+\frac12\frac{\dot{F}}{F}\right)^2}
\left[\left(\frac{Hw_{\varphi}}{\dot{\phi}}\right)^{\cdot 2}
-\left(\frac{\partial}{a}\frac{Hw_{\varphi}}{\dot{\phi}}\right)^2\right]
\,,
\label{nmaction:2:wvarphiwvarphi:GI}
\end{align}
The first order equations for motion for the graviton
and the gauge invariant curvature perturbation are
\begin{align}
\frac{\delta S^{(2)}}{\delta \gamma_{ij}}=&0=
\frac{a^3F}{4}\left(-\frac{1}{a^{3}F}\left(a^{3}F\dot{\gamma}_{ij}\right)^{\cdot}
+\frac{\nabla^2}{a^2}\gamma_{ij}\right)
\nonumber\\
\frac{\delta S^{(2)}}{\delta w_{\zeta}}=&0=
-\left(a^{3}\frac{\dot{\phi}^2+\frac{3}{2}\frac{\dot{F}^2}{F}}
{\left(H+\frac{1}{2}\frac{\dot{F}}{F}\right)^2}\dot{w}_{\zeta}\right)^{\cdot}
+a^3\frac{\dot{\phi}^2+\frac{3}{2}\frac{\dot{F}^2}{F}}
{\left(H+\frac{1}{2}\frac{\dot{F}}{F}\right)^2}
\frac{\nabla^2}{a^2}w_{\zeta}
\,.
\label{firstorderequationsofmotion}
\end{align}
The first order equation of motion for the gauge invariant
field perturbation is related to that of the curvature perturbation
as
\begin{equation}
\frac{\delta S^{(2)}}{\delta w_{\varphi}}=0=
\frac{-H}{\dot{\phi}}
\left.\frac{\delta S^{(2)}}{\delta w_{\zeta}}\right|_{w_{\zeta}=-\frac{Hw_{\varphi}}{\dot{\phi}}}
\,.
\label{firstorderequationsofmotion:wvarphi}
\end{equation}
This is of course true since at first order the
gauge invariant curvature perturbation and
field perturbation are related via a rescaling
by background quantities
\begin{equation}
w_{\zeta}=-\frac{H}{\dot{\phi}}w_{\varphi}
\,.
\label{relationfirstorderscalarperturbations}
\end{equation}
The fact that the linear gauge invariant variables on
different hypersurfaces are related by time dependent
rescalings, plus the fact that the graviton decouples
and is gauge invariant by itself, makes it relatively
easy to find scalar and graviton 2-point functions on different
hypersurfaces. At higher order this is no longer the case,
and we shall discuss this next.

\subsection{Third order gauge invariant action}
\label{sec: Third order gauge invariant action}

We continue by finding the gauge invariant vertices
for scalar and graviton interactions in the Jordan frame.
Due to the non-linear nature of general relativity
the second order perturbations \eqref{perturbations}
transform non-linearly under gauge transformations.
As such, the gauge invariant scalar and tensor
perturbations receive quadratic contributions.
This implies that some of the vertices for
scalar and graviton interactions that naively
appear in the third order action after inserting
the perturbations \eqref{perturbations} in \eqref{ADMactionnonminimalEij},
are actually non-physical. When the naive third order action
is expressed in terms of second order gauge invariant
variables, some of the cubic vertices are absorbed
into the quadratic action for these variables.
This greatly complicates the construction of the third
order action for scalar and graviton interactions.
However, there is a systematic way to isolate the
physical cubic vertices and at the same time find
the second order gauge invariant scalar and tensor
perturbation. The procedure was outlined in Ref. \cite{Prokopec:2012ug}
where the cubic vertices for scalar interactions were found.
Here we apply the procedure to the cubic action for
scalar-graviton interactions.

\subsubsection{Scalar-scalar-graviton vertices on uniform field hypersurfaces}
\label{sec: Scalar-scalar-graviton vertices on uniform field hypersurfaces}
The naive scalar-scalar-graviton vertices are found
by collecting those third order terms from the action
\eqref{ADMactionnonminimalEij} expanded up to third
order in perturbations \eqref{perturbations}.
This gives
\begin{align}
S[ssg]=\frac12\int d^{3}xdt \bar{N} a^{3}
\Biggl\{&
4\left[F\zeta+F'\varphi+F n \right]\gamma_{ij}\frac{\partial_i\partial_j\zeta}{a^2}
+2F\gamma_{ij}\frac{\partial_i\zeta}{a}\frac{\partial_j\zeta}{a}
\nonumber\\
&+\left[-3F\zeta-F'\varphi+nF\right]\dot{\gamma}_{ij}\frac{\partial_i\partial_j s}{a^2}
+F\frac{\partial_i\partial_js}{a^2}\frac{\partial_k \gamma_{ij}}{a}\frac{\partial_k s}{a}
\nonumber\\
&-2\gamma_{ij}\frac{\partial_i n}{a}F'\frac{\partial_j\varphi}{a}
+\gamma_{ij}\frac{\partial_i \varphi}{a}\frac{\partial_j\varphi}{a}
\Biggr\}
\,.
\label{nmaction:3:ssg}
\end{align}
Note that $s$ and $n$ are linear in scalar perturbations
according to \eqref{lapsesolution:nonminimal} and \eqref{shiftsolution:nonminimal:b}.
Gauge invariance is not manifest in the action \eqref{nmaction:3:ssg},
because the scalar perturbations $\zeta$ and $\varphi$ are not linearly
gauge invariant. The curvature perturbation on uniform field hypersurfaces,
$w_{\zeta}$, on the other hand is gauge invariant to first order.
So let us find the gauge invariant cubic vertices for scalar-scalar-graviton
interactions by inserting into the action above
\begin{equation}
\zeta=w_{\zeta}+\frac{H}{\dot{\phi}}\varphi
\,.
\label{GIscalar:replacement zeta}
\end{equation}
This substitution separates the action \eqref{nmaction:3:ssg}
into a manifestly gauge invariant part with $w_{\zeta}^2\gamma$
vertices, and a gauge dependent part with $\varphi^2\gamma$
and $w_{\zeta}\varphi\gamma$ vertices. Schematically
\begin{equation}
S[ssg]=S_{\rm GI}[w_{\zeta}^2\gamma]+S_{\rm GD}[\varphi^2\gamma+w_{\zeta}\varphi\gamma]
\label{separationaction:ssg}
\,.
\end{equation}
The gauge invariant vertices are
\begin{align}
S_{\rm GI}[w_{\zeta}^2\gamma]=\frac12\int d^{3}xdt \bar{N} a^{3}F
\Biggl\{&
4\left[w_{\zeta}+n_{w_{\zeta}}\right]\gamma_{ij}\frac{\partial_i\partial_jw_{\zeta}}{a^2}
+2\gamma_{ij}\frac{\partial_iw_{\zeta}}{a}\frac{\partial_jw_{\zeta}}{a}
\nonumber\\
&- \left[3w_{\zeta}-n_{w_{\zeta}}\right]\dot{\gamma}_{ij}\frac{\partial_i\partial_j s_{w_{\zeta}}}{a^2}
+\frac{\partial_i\partial_js_{w_{\zeta}}}{a^2}\frac{\partial_k \gamma_{ij}}{a}\frac{\partial_k s_{w_{\zeta}}}{a}
\Biggr\}
\,,
\label{nmaction:3:scalarscalargraviton:GI}
\end{align}
where for convenience we have separated the first order solutions
of the constraint equations in a gauge invariant and gauge
dependent part,
\begin{align}
n=& \frac{1}{H+\frac{1}{2}\frac{\dot{F}}{F}}\dot{w}_{\zeta}
+\left(\frac{\varphi}{\dot{\phi}}\right)^{\cdot}
\equiv  n_{w_{\zeta}}+\left(\frac{\varphi}{\dot{\phi}}\right)^{\cdot}
\nonumber\\
\frac{\nabla^2s}{a^2}
=&\frac{\nabla^2}{a^2}\left(
\frac{-1}{H+\frac{1}{2}\frac{\dot{F}}{F}}w_{\zeta}
+\chi\right)
-\frac{\nabla^2}{a^2}\frac{\varphi}{\dot{\phi}}
\equiv \frac{\nabla^2}{a^2}s_{w_{\zeta}}
-\frac{\nabla^2}{a^2}\frac{\varphi}{\dot{\phi}}
\,.
\label{lapsesolution:splitwzeta:nonminimal}
\end{align}
Note from Eq. \eqref{shiftsolution:nonminimal:b} that
the $\chi$ part of the first order solution is gauge
invariant by itself.\\
Although we did not write the gauge dependent part of the third order
action $S_{\rm GD}[\varphi^2\gamma+w_{\zeta}\varphi\gamma]$
explicitly, it is easy to see that it is nonzero.
This seems to pose a major problem, since it suggests
that gauge invariance is broken at the perturbative level.
However, we should remind ourselves that in the third order
action we deal with second order perturbations. Likewise,
these perturbations transform to second order under gauge
transformations. As a consequence, the first order gauge invariant
curvature perturbation $w_{\zeta}$ and graviton $\gamma_{ij}$
are no longer gauge invariant. Schematically we can write
these second order gauge transformations as (see Refs. \cite{Rigopoulos:2011eq}
\cite{Prokopec:2012ug})
\begin{align}
w_{\zeta}&\rightarrow w_{\zeta}+\Delta^{\xi}_{2}w_{\zeta}
\nonumber\\
\gamma_{ij}&\rightarrow \gamma_{ij} +(\Delta^{\xi}_{2}\gamma)_{ij}
\,.
\end{align}
Under such gauge transformations the second order actions
for $w_{\zeta}$ and $\gamma_{ij}$ transform as
\begin{align}
S^{(2)}[\gamma^2]&\rightarrow S^{(2)}[\gamma^2]
+\frac12 \int d^3xdt\bar{N}a^3\left\{
\frac{2}{a^3} \frac{\delta S^{(2)}}{\delta \gamma_{ij}}
(\Delta^{\xi}_{2} \gamma)_{ij}
\right\}
\nonumber\\
S^{(2)}[w_{\zeta}^2]&\rightarrow S^{(2)}[w_{\zeta}^2]
+\frac12 \int d^3xdt\bar{N}a^3\left\{
\frac{2}{a^3} \frac{\delta S^{(2)}}{\delta w_{\zeta}}
\Delta^{\xi}_{2}w_{\zeta}
\right\}
\,,
\label{gaugetransformationsactionwzetagamma}
\end{align}
up to total derivatives. This second order gauge transformation
of the quadratic action can only be balanced by gauge dependent
terms in the third order action which are proportional to the
equation of motion. So in fact, in order for the third order
action to be gauge invariant under second order transformations,
we expect to have a gauge dependent part in the third order
action after the substitution \eqref{GIscalar:replacement zeta}
into \eqref{ADMactionnonminimalEij}. This gauge dependent
part must be proportional to the linear equations of motion.
Indeed, after many partial integrations we find that
\begin{align}
S_{\rm GD}[\varphi^2\gamma+w_{\zeta}\varphi\gamma]=\frac12\int d^3xdt \bar{N} a^{3}
\Biggl\{&
\frac{2}{a^3}\frac{\delta S^{(2)}}{\delta \gamma_{ij}}
\left[\frac{\partial_i\varphi}{a\dot{\phi}}
\frac{\partial_j\varphi}{a\dot{\phi}}
-\left(\frac{\partial_i\varphi}{a\dot{\phi}}\frac{\partial_j s_{w_{\zeta}}}{a}
+\frac{\partial_j\varphi}{a\dot{\phi}}\frac{\partial_i s_{w_{\zeta}}}{a}
\right)\right]
\nonumber\\
&+\frac{2}{a^3} \frac{\delta S^{(2)}}{\delta w_{\zeta}}
\left[\frac{\partial_i\partial_j}{\nabla^2}
\left(\frac{1}{4}\frac{\varphi}{\dot{\phi}}\dot{\gamma}_{ij}\right)
\right]
\Biggr\}
\,.
\label{nmaction:3:scalarscalargraviton:GD:wvarphigamma:rewritten}
\end{align}
The complete action to third order in perturbations
can now be written in a manifestly gauge invariant
way by defining second order gauge invariant variables,
\begin{align}
\tilde{\gamma}_{\zeta,ij}=&\gamma_{ij}
+\frac{\partial_i\varphi}{a\dot{\phi}}
\frac{\partial_j\varphi}{a\dot{\phi}}
-\left(\frac{\partial_i\varphi}{a\dot{\phi}}\frac{\partial_j s_{w_{\zeta}}}{a}
+\frac{\partial_j\varphi}{a\dot{\phi}}\frac{\partial_i s_{w_{\zeta}}}{a}
\right)
+\mathcal{O}(\gamma \varphi)
\label{gammaijzeta:GI2ndorder}
\\
W_{\zeta}=&w_{\zeta}
+\frac{\partial_i\partial_j}{\nabla^2}
\left(\frac{1}{4}\frac{\varphi}{\dot{\phi}}\dot{\gamma}_{ij}\right)
+\mathcal{O}(\varphi^2)
\label{Wzeta:GI2ndorder}
\,.
\end{align}
The first variable is the second order gauge invariant
tensor perturbation, and may be called the \textit{graviton
on uniform field hypersurfaces}, since it reduces to $\gamma_{ij}$
in the gauge $\varphi=0$. Similarly, $W_{\zeta}$ is the second
order curvature perturbation on uniform field hypersurfaces.
The manifestly gauge invariant quadratic actions
for these perturbations have precisely the
same form as Eqs. \eqref{nmaction:2:gammagamma} and
\eqref{nmaction:2:wzetawzeta:GI}, \textit{i.e.}
\begin{align}
S^{(2)}[\tilde{\gamma}_{\zeta}^2]&=\frac12\int d^3xdt \bar{N} a^3\frac{F}{4}
\left[\dot{\tilde{\gamma}}_{\zeta,ij}\dot{\tilde{\gamma}}_{\zeta,ij}
-\left(\frac{\partial \tilde{\gamma}_{\zeta,ij}}{a}\right)^2\right]
\nonumber\\
S^{(2)}[W_{\zeta}^2]&=\frac12\int d^{3}xdt \bar{N} a^{3}
\frac{\dot{\phi}^2+\frac{3}{2}\frac{\dot{F}^2}{F}}
{\left(H+\frac12\frac{\dot{F}}{F}\right)^2}
\left[\dot{W}_{\zeta}^2-\left(\frac{\partial W_{\zeta}}{a}\right)^2\right]
\,.
\label{nmaction:2:WzetaWzeta:GI}
\end{align}
The vertices for scalar-scalar-graviton interactions
can be found immediately from Eq. \eqref{nmaction:3:scalarscalargraviton:GI}
by replacing the first order gauge invariant variables
by their second order generalizations (which does not change
anything for the cubic vertices). Thus we get
\begin{align}
S[W_{\zeta}^2\tilde{\gamma}_{\zeta}]=\frac12\int d^{3}xdt \bar{N} a^{3}F
\Biggl\{&
4\left[W_{\zeta}+n_{W_{\zeta}}\right]
\tilde{\gamma}_{\zeta,ij}\frac{\partial_i\partial_jW_{\zeta}}{a^2}
+2\tilde{\gamma}_{\zeta,ij}\frac{\partial_iW_{\zeta}}{a}\frac{\partial_jW_{\zeta}}{a}
\nonumber\\
&- \left[3W_{\zeta}-n_{W_{\zeta}}\right]
\dot{\tilde{\gamma}}_{\zeta,ij}\frac{\partial_i\partial_j W_{w_{\zeta}}}{a^2}
+\frac{\partial_i\partial_js_{W_{\zeta}}}{a^2}
\frac{\partial_k \tilde{\gamma}_{\zeta,ij}}{a}\frac{\partial_k s_{W_{\zeta}}}{a}
\Biggr\}
\,.
\label{nmaction:3:ssg:GI:Wzeta}
\end{align}
Note that the gauge invariant variables can still contain
other terms at second order, as can be seen from Eqs. \eqref{Wzeta:GI2ndorder}.
The $\mathcal{O}(\gamma\varphi)$ terms in the definition
for $\tilde{\gamma}_{\zeta}$ can only originate
from the scalar-graviton-graviton action at third order,
which is precisely what we shall discuss in the next section.
The $\mathcal{O}(\varphi^2)$ terms in the definition
for $W_{\zeta}$ originate from the cubic scalar interactions.
They were discussed (in the long wavelength limit) in
Refs. \cite{Rigopoulos:2011eq} and \cite{Prokopec:2012ug}.

In conclusion, we have presented and applied
a systematic method to find the manifestly gauge
invariant action at third order. The trick is to
simply insert the first order gauge invariant variable
\eqref{GIscalar:replacement zeta} into the naive cubic vertices .
Not only does this method give the physical cubic vertices,
but it also gives the correct form of the second order
gauge invariant variables. Usually the second order gauge
invariant variables are found by fixing the second order
gauge transformation of the perturbations (see \textit{e.g.}
\cite{Malik:2003mv,Malik:2008im}). Here no explicit second order
gauge transformations were necessary \footnote{Of course,
one could check that the second order gauge invariant variables
which follow from the action procedure agree with those
constructed via the gauge transformation method as in
\cite{Malik:2008im}.}. The only ingredient
that we used is the manifest general covariance of the complete
action \eqref{nonminimalaction}, and the notion that
a manifestly gauge invariant can be written down order
by order in perturbation theory.

The method can also be generalized to higher order perturbation theory.
Take the manifestly gauge invariant action for
second order perturbations, Eqs. \eqref{nmaction:2:WzetaWzeta:GI}
and \eqref{nmaction:3:ssg:GI:Wzeta}.
The action \eqref{ADMactionnonminimalEij} expanded to fourth order
in perturbation theory contains again $\zeta$, $\varphi$ and $\gamma_{ij}$.
Now replace $\zeta=W_{\zeta}+H\varphi/\dot{\phi}$
and $\gamma_{ij}=\tilde{\gamma}_{\zeta,ij}$. The fourth order
action separates in a gauge invariant part with second order
gauge invariant variables $W_{\zeta}$ and $\tilde{\gamma}_{\zeta,ij}$,
and a gauge dependent part. The gauge dependent part must
be must be proportional to the linear equations of motion,
\textit{and} to $\delta S^{3} / \delta w_{\zeta}$ and
$\delta S^{3} / \delta \gamma_{ij}$. The first part must be there
to balance the third order gauge transformation of
the quadratic action, and can be absorbed into the definition
of third order gauge invariant variables. The second part must be
there to balance the second order gauge transformation of the
cubic action, and can be absorbed into the definition of second
order gauge invariant variables in the cubic action. Note that
we replaced $w_{\zeta}\rightarrow W_{\zeta}$ in the cubic vertices,
which is only allowed when we consider the action up to third
order in perturbations. One could go to even higher order in perturbation
theory, where in general the gauge dependent part of the $n$th order
action can be absorbed into lower order actions by defining
gauge invariant variables at $n-1$th order.

\subsubsection{Scalar-graviton-graviton vertices on uniform field hypersurfaces}
\label{sec: Scalar-graviton-graviton vertices on uniform field hypersurfaces}

So far we have only discussed the scalar-scalar-graviton vertices.
The gauge invariant scalar-graviton-graviton and pure graviton vertices
can be computed using the same method as described in the previous section.
We start with the naive scalar-graviton-graviton vertices from
the expansion of \eqref{ADMactionnonminimalEij}
\begin{align}
S[sgg]=\frac12\int d^{3}xdt \bar{N} a^{3}
\Biggl\{&
-2F\gamma_{ik}\gamma_{kj}\frac{\partial_i\partial_j\zeta}{a^2}
-2F\gamma_{ik}\frac{\partial_i\gamma_{jk}}{a}\frac{\partial_j\zeta}{a}
\nonumber\\
&+\left(3F\zeta+F'\varphi\right)\frac14 \dot{\gamma}_{kl}\dot{\gamma}_{kl}
-\left(F\zeta+F'\varphi\right)
\frac14\frac{\partial_i\gamma_{kl}}{a}\frac{\partial_i\gamma_{kl}}{a}
\nonumber\\
&-n F \left(\frac14 \dot{\gamma}_{kl}\dot{\gamma}_{kl}
+\frac14\frac{\partial_i\gamma_{kl}}{a}\frac{\partial_i\gamma_{kl}}{a}\right)
-\frac12 F \dot{\gamma}_{kl}\frac{\partial_i\gamma_{kl}}{a}\frac{\partial_is}{a}
\Biggr\}
\,.
\label{nmaction:3:scalargravitongraviton}
\end{align}
Again we separate the cubic terms in manifestly gauge invariant
vertices composed of $w_{\zeta}$ and $\gamma_{ij}$, plus
gauge non-invariant terms, by using \eqref{GIscalar:replacement zeta}.
The gauge invariant vertices are
\begin{align}
S_{\rm GI}[w_{\zeta}\gamma^2]=\frac12\int d^{3}xdt \bar{N} a^{3}F
\Biggl\{&
-2\gamma_{ik}\gamma_{kj}\frac{\partial_i\partial_jw_{\zeta}}{a^2}
-2\gamma_{ik}\frac{\partial_i\gamma_{jk}}{a}\frac{\partial_jw_{\zeta}}{a}
-\frac12 \dot{\gamma}_{kl}\frac{\partial_i\gamma_{kl}}{a}\frac{\partial_is_{w_{\zeta}}}{a}
\nonumber\\
&+(3w_{\zeta}-n_{w_{\zeta}})\frac14 \dot{\gamma}_{kl}\dot{\gamma}_{kl}
-(w_{\zeta}+n_{w_{\zeta}})
\frac14\frac{\partial_i\gamma_{kl}}{a}\frac{\partial_i\gamma_{kl}}{a}
\Biggr\}
\,,
\label{nmaction:3:scalargravitongraviton:GI}
\end{align}
and the gauge dependent part of the action \eqref{nmaction:3:scalargravitongraviton}
(with $\varphi\gamma^2$ terms)
becomes after several partial integrations
\begin{align}
S_{\rm GD}[\varphi\gamma^2]=\frac12\int d^{3}xdt \bar{N} a^{3}
\Biggl\{&-\frac{\varphi}{\dot{\phi}}\gamma_{ij}
\frac{2}{a^3}\frac{\delta S^{(2)}}{\delta \gamma_{ij}}
\Biggr\}
\,,
\label{nmaction:3:scalargravitongraviton:GD:rewritten}
\end{align}
up to boundary terms. Hence, this gauge dependent part of the
scalar-graviton-graviton action may be absorbed by a redefinition
of $\gamma_{ij}$. Considering the previously defined
second order graviton on uniform field hypersurfaces $\tilde{\gamma}_{\zeta,ij}$,
this precisely describes the $\mathcal{O}(\varphi\gamma)$ contribution.
Thus, the complete gauge invariant graviton to second order is
\begin{align}
\tilde{\gamma}_{\zeta,ij}=&\gamma_{ij}
+\frac{\partial_i\varphi}{a\dot{\phi}}
\frac{\partial_j\varphi}{a\dot{\phi}}
-\left(\frac{\partial_i\varphi}{a\dot{\phi}}\frac{\partial_j s_{w_{\zeta}}}{a}
+\frac{\partial_j\varphi}{a\dot{\phi}}\frac{\partial_i s_{w_{\zeta}}}{a}
\right)
-\frac{\varphi}{\dot{\phi}}\dot{\gamma}_{ij}
\,.
\label{gammaijzeta:GI2ndorder:complete}
\end{align}
The manifestly gauge invariant cubic vertices
for scalar-graviton-graviton interactions are
then
\begin{align}
S[W_{\zeta}\tilde{\gamma}_{\zeta}^2]=\frac12\int d^{3}xdt \bar{N} a^{3}F
\Biggl\{&
-2\tilde{\gamma}_{\zeta,ik}\tilde{\gamma}_{\zeta,kj}\frac{\partial_i\partial_jW_{\zeta}}{a^2}
-2\tilde{\gamma}_{\zeta,ik}\frac{\partial_i\tilde{\gamma}_{\zeta,jk}}{a}\frac{\partial_jW_{\zeta}}{a}
-\frac12 \dot{\tilde{\gamma}}_{\zeta,kl}\frac{\partial_i\tilde{\gamma}_{\zeta,kl}}{a}\frac{\partial_is_{W_{\zeta}}}{a}
\nonumber\\
&+(3W_{\zeta}-n_{W_{\zeta}})\frac14 \dot{\tilde{\gamma}}_{\zeta,kl}\dot{\tilde{\gamma}}_{\zeta,kl}
-(W_{\zeta}+n_{W_{\zeta}})
\frac14\frac{\partial_i\tilde{\gamma}_{\zeta,kl}}{a}\frac{\partial_i\tilde{\gamma}_{\zeta,kl}}{a}
\Biggr\}
\,.
\label{nmaction:3:sgg:GI:Wzeta}
\end{align}

\subsubsection{Pure graviton vertices on uniform field hypersurfaces}
\label{sec: Pure graviton vertices on uniform field hypersurfaces}

Since the second order tensor perturbation $\gamma_{ij}$ is gauge invariant
to first order, the pure graviton vertices are automatically gauge invariant.
Thus the $\gamma^3$ found after an expansion of \eqref{ADMactionnonminimalEij}
are the physical vertices. In the cubic vertices we can replace
$\gamma_{ij}$ by its second order gauge invariant generalization,
$\tilde{\gamma}_{\zeta,ij}$, which gives the manifestly gauge invariant vertices
\begin{align}
S[\tilde{\gamma}_{\zeta}^3]=\frac12\int d^3xdt \bar{N} a^3 F
\Biggl\{&
\frac14 \tilde{\gamma}_{\zeta,ij}
\frac{\partial_i\tilde{\gamma}_{\zeta,kl}}{a}
\frac{\partial_j\tilde{\gamma}_{\zeta,kl}}{a}
+\frac14 \tilde{\gamma}_{\zeta,kl}
\frac{\partial_i\tilde{\gamma}_{\zeta,kj}}{a}
\frac{\partial_j\tilde{\gamma}_{\zeta,il}}{a}
-\frac14 \tilde{\gamma}_{\zeta,ik}
\frac{\partial_i\tilde{\gamma}_{\zeta,jl}}{a}
\frac{\partial_j\tilde{\gamma}_{\zeta,kl}}{a}
\Biggr\}
\,.
\label{nmaction:3:ggg:GI:Wzeta}
\end{align}

\subsubsection{Scalar-graviton interactions on uniform curvature hypersurfaces}
\label{sec: Scalar-graviton interactions on uniform curvature hypersurfaces}

In the previous sections we we constructed in a systematic
the cubic gauge invariant action for the second order
graviton and curvature perturbation on uniform field hypersurfaces.
Similarly, we can find the gauge invariant action
for scalar field and tensor perturbations on uniform
curvature hypersurfaces. The starting point is again the
action \eqref{ADMactionnonminimalEij}, which expanded
to third order for scalar-graviton interactions gives
Eqs. \eqref{nmaction:3:ssg} and \eqref{nmaction:3:scalargravitongraviton}.
Instead of replacing $\zeta$ by its linearly gauge invariant
variable $w_{\zeta}$, we now make use of Eq. \eqref{wvarphi:firstorder}
to replace
\begin{equation}
\varphi=w_{\varphi}+\frac{\dot{\phi}}{H}\zeta
\,.
\label{GIscalar:replacement varphi}
\end{equation}
This separates the cubic vertices into a manifestly
gauge invariant part involving linearly gauge invariant
scalar perturbation $w_{\varphi}$ and tensor $\gamma_{ij}$,
plus gauge dependent vertices that involve the
gauge dependent variable $\zeta$. Analogous
to the previous sections, the gauge dependent part
of the cubic action can be brought to a form
which is proportional to the linear equations
of motion for $w_{\varphi}$ and $\gamma_{ij}$. These
terms may be absorbed into the quadratic action
by defining second order variables
\begin{align}
\tilde{\gamma}_{\varphi,ij}=&\gamma_{ij}-\frac{\zeta}{H}\dot{\gamma}_{ij}
+\frac{\partial_i\zeta}{aH}
\frac{\partial_j\zeta}{aH}
-\left(\frac{\partial_i\zeta}{aH}\frac{\partial_j s_{w_{\varphi}}}{a}
+\frac{\partial_j\zeta}{aH}\frac{\partial_i s_{w_{\varphi}}}{a}
\right)
\nonumber\\
W_{\varphi}=&w_{\varphi}
-\frac{\partial_i\partial_j}{\nabla^2}\left(\frac{1}{4}\frac{\dot{\phi}\zeta}{H^2}\dot{\gamma}_{ij}\right)
+\mathcal{O}(\zeta^2)
\,,
\label{Wvarphi:gammavarphi:GI2ndorder}
\end{align}
where we have separated the first order constraint
solutions \eqref{lapsesolution:nonminimal} and
\eqref{shiftsolution:nonminimal:b} into a gauge
invariant and gauge dependent part
\begin{align}
n=& -\frac{\dot{H}}{H\dot{\phi}}w_{\varphi}
+\frac{\frac12\frac{\dot{F}}{F}}{H+\frac{1}{2}\frac{\dot{F}}{F}}
\left(\frac{Hw_{\varphi}}{\dot{\phi}}\right)^{\cdot}
+\left(\frac{\zeta}{H}\right)^{\cdot}
\equiv  n_{w_{\varphi}}+\left(\frac{\zeta}{H}\right)^{\cdot}
\nonumber\\
\frac{\nabla^2s}{a^2}
=&\frac{\nabla^2}{a^2}\left(
\frac{-\frac12\frac{\dot{F}}{F}}{H+\frac{1}{2}\frac{\dot{F}}{F}}
\frac{w_{\varphi}}{\dot{\phi}}
+\chi\right)
-\frac{\nabla^2}{a^2}\frac{\zeta}{H}
\equiv \frac{\nabla^2}{a^2}s_{w_{\varphi}}
-\frac{\nabla^2}{a^2}\frac{\zeta}{H}
\,.
\label{lapsesolution:splitwvarphi:nonminimal}
\end{align}
In the gauge $\zeta=0$ the perturbations $\tilde{\gamma}_{\varphi,ij}$
and $W_{\varphi}$ reduce to the graviton $\gamma_{ij}$ and
field perturbation $\varphi$. Therefore, they are called
the second order gauge invariant \textit{graviton on
uniform curvature hypersurfaces} and \textit{field
perturbation on uniform curvature hypersurfaces}, respectively.
The $\mathcal{O}(\zeta^2)$ terms in the second order
perturbation $W_{\varphi}$ follow from the pure scalar
interactions, which have been discussed (in part) in
\cite{Rigopoulos:2011eq} and \cite{Prokopec:2012ug}.
The quadratic action for the perturbations
\eqref{Wvarphi:gammavarphi:GI2ndorder} is
\begin{align}
S^{(2)}[\tilde{\gamma}_{\varphi}^2]&=\frac12\int d^3xdt \bar{N} a^3\frac{F}{4}
\left[\dot{\tilde{\gamma}}_{\varphi,ij}\dot{\tilde{\gamma}}_{\varphi,ij}
-\left(\frac{\partial \tilde{\gamma}_{\varphi,ij}}{a}\right)^2\right]
\nonumber\\
S^{(2)}[W_{\varphi}^2]&=\frac12\int d^{3}xdt \bar{N} a^{3}
\frac{\dot{\phi}^2+\frac{3}{2}\frac{\dot{F}^2}{F}}
{\left(H+\frac12\frac{\dot{F}}{F}\right)^2}
\left[\left(\frac{H W_{\varphi}}{\dot{\phi}}\right)^{\cdot 2}
-\left(\frac{\partial}{a}\frac{H W_{\varphi}}{\dot{\phi}}\right)^2\right]
\,.
\label{nmaction:2:WvarphiWvarphi:GI}
\end{align}
The interaction vertices are found by replacing $w_{\varphi}$
and $\gamma_{ij}$ in the gauge invariant part of the third
order action by their second order generalizations.
This gives the scalar-scalar-graviton vertices
\begin{align}
S[W_{\varphi}^2\tilde{\gamma}_{\varphi}]=\frac12\int d^{3}xdt \bar{N} a^{3}
\Biggl\{&
\left(n_{W_{\varphi}}F-\dot{F}\frac{W_{\varphi}}{\dot{\phi}}\right)
\dot{\tilde{\gamma}}_{\varphi,ij}\frac{\partial_i\partial_j s_{W_{\varphi}}}{a^2}
-2\dot{F}\tilde{\gamma}_{\varphi,ij}\frac{\partial_i n_{W_{\varphi}}}{a}
\frac{\partial_j}{a}\frac{W_{\varphi}}{\dot{\phi}}
\nonumber\\
&
+\tilde{\gamma}_{\varphi,ij}\frac{\partial_i W_{\varphi}}{a}\frac{\partial_jW_{\varphi}}{a}
+F\frac{\partial_i\partial_js_{W_{\varphi}}}{a^2}
\frac{\partial_k \tilde{\gamma}_{\varphi,ij}}{a}\frac{\partial_k s_{W_{\varphi}}}{a}
\Biggr\}
\,,
\label{nmaction:3:ssg:GI:Wvarphi}
\end{align}
and the scalar-graviton-graviton vertices
\begin{align}
S[W_{\varphi}\tilde{\gamma}_{\varphi}^2]=\frac12\int d^{3}xdt \bar{N} a^{3}
\Biggl\{&
\dot{F}\frac{W_{\varphi}}{\dot{\phi}}\left(\frac14 \dot{\tilde{\gamma}}_{\varphi,kl}\dot{\tilde{\gamma}}_{\varphi,kl}
-\frac14\frac{\partial_i\tilde{\gamma}_{\varphi,kl}}{a}\frac{\partial_i\tilde{\gamma}_{\varphi,kl}}{a}\right)
\nonumber\\
&-n_{W_{\varphi}} F \left(\frac14 \dot{\tilde{\gamma}}_{\varphi,kl}\dot{\tilde{\gamma}}_{\varphi,kl}
+\frac14\frac{\partial_i\tilde{\gamma}_{\varphi,kl}}{a}\frac{\partial_i\tilde{\gamma}_{\varphi,kl}}{a}\right)
-\frac12 \dot{\tilde{\gamma}}_{\varphi,kl}\frac{\partial_i\tilde{\gamma}_{\varphi,kl}}{a}\frac{\partial_is_{W_{\varphi}}}{a}
\Biggr\}
\,,
\label{nmaction:3:sgg:GI:Wvarphi}
\end{align}
and finally the pure graviton vertices
\begin{align}
S[\tilde{\gamma}_{\varphi}^3]=\frac12\int d^3xdt \bar{N} a^3 F
\Biggl\{&
\frac14 \tilde{\gamma}_{\varphi,ij}
\frac{\partial_i\tilde{\gamma}_{\varphi,kl}}{a}
\frac{\partial_j\tilde{\gamma}_{\varphi,kl}}{a}
+\frac14 \tilde{\gamma}_{\varphi,kl}
\frac{\partial_i\tilde{\gamma}_{\varphi,kj}}{a}
\frac{\partial_j\tilde{\gamma}_{\varphi,il}}{a}
-\frac14 \tilde{\gamma}_{\varphi,ik}
\frac{\partial_i\tilde{\gamma}_{\varphi,jl}}{a}
\frac{\partial_j\tilde{\gamma}_{\varphi,kl}}{a}
\Biggr\}
\,.
\label{nmaction:3:ggg:GI:Wvarphi}
\end{align}
In these interaction actions the $n_{W_{\varphi}}$ and
$s_{W_{\varphi}}$ are obtained by replacing
$w_{\varphi}\rightarrow W_{\varphi}$ in
Eqs. \eqref{lapsesolution:splitwvarphi:nonminimal}.

\subsection{Uniqueness of the scalar-graviton action}
\label{sec: Uniqueness of the scalar-graviton action}
We have now computed two gauge invariant actions for cubic scalar
and graviton interactions. Eqs. \eqref{nmaction:3:ssg:GI:Wzeta},
\eqref{nmaction:3:sgg:GI:Wzeta} and \eqref{nmaction:3:ggg:GI:Wzeta}
describe the interactions for the curvature perturbation
and graviton on uniform field hypersurfaces,
$W_{\zeta}$ and $\tilde{\gamma}_{\zeta,ij}$.
Eqs. \eqref{nmaction:3:ssg:GI:Wvarphi},
\eqref{nmaction:3:sgg:GI:Wvarphi} and \eqref{nmaction:3:ggg:GI:Wvarphi}
describe the interactions for the field perturbation
and graviton on uniform curvature hypersurfaces,
$W_{\varphi}$ and $\tilde{\gamma}_{\varphi,ij}$.
If we compare the gauge invariant vertices on
different hypersurfaces, we see that they are not the
same. They are however related in a relatively simple way.
We can show this by writing the cubic actions on different
hypersurfaces in the same form. For example,
the scalar-scalar-graviton interactions Eqs.
\eqref{nmaction:3:ssg:GI:Wzeta} and \eqref{nmaction:3:ssg:GI:Wvarphi}
may be partially integrated in order to find,
up to boundary terms,
\begin{align}
S[W_{\zeta}^2\tilde{\gamma}_{\zeta}]=\frac12\int d^{3}xdt \bar{N} a^{3}
&\Biggl\{
-\frac{1}{2}\frac{\dot{\phi}^2+\frac{3}{2}\frac{\dot{F}^2}{F}}
{\left(H+\frac{1}{2}\frac{\dot{F}}{F}\right)^2}
W_{\zeta}
\dot{\tilde{\gamma}}_{\zeta,ij}\frac{\partial_i\partial_j \chi}{a^2}
+\frac{\dot{\phi}^2+\frac{3}{2}\frac{\dot{F}^2}{F}}{\left(H+\frac{1}{D-2}\frac{\dot{F}}{F}\right)^2}
\tilde{\gamma}_{\zeta,ij}
\frac{\partial_i W_{\zeta}}{a}\frac{\partial_jW_{\zeta}}{a}
\nonumber\\
&
+F\frac{\partial_i\partial_j\chi}{a^2}
\frac{\partial_k \tilde{\gamma}_{\zeta,ij}}{a}\frac{\partial_k \chi}{a}
+\frac{\partial_i\partial_j}{\nabla^2}\left[
\frac{\dot{\tilde{\gamma}}_{\zeta,ij}}{4}
\frac{W_{\zeta}}{H+\frac{1}{2}\frac{\dot{F}}{F}}
\right]
\frac{2}{a^3}\frac{\delta S^{(2)}}{\delta W_{\zeta}}
\nonumber\\
&
+\frac{1}{H+\frac{1}{2}\frac{\dot{F}}{F}}
\left[\frac{1}{H+\frac{1}{2}\frac{\dot{F}}{F}}
\frac{\partial_i W_{\zeta}}{a}\frac{\partial_j W_{\zeta}}{a}
-\left(\frac{\partial_i W_{\zeta}}{a}\frac{\partial_j \chi}{a}
+\frac{\partial_j W_{\zeta}}{a}\frac{\partial_i \chi}{a}
\right)
\right]\frac{2}{a^3}\frac{\delta S^{(2)}}{\delta \tilde{\gamma}_{\zeta,ij}}
\Biggr\}
\,,
\label{nmaction:3:ssg:Wzeta:2}
\end{align}
and
\begin{align}
S[W_{\varphi}^2\tilde{\gamma}_{\varphi}]=\frac12\int d^{3}xdt \bar{N} a^{3}
&\Biggl\{
\frac{1}{2}\frac{\dot{\phi}^2+\frac{3}{2}\frac{\dot{F}^2}{F}}
{\left(H+\frac{1}{2}\frac{\dot{F}}{F}\right)^2}
\frac{HW_{\varphi}}{\dot{\phi}}
\dot{\tilde{\gamma}}_{\varphi,ij}\frac{\partial_i\partial_j \chi}{a^2}
+\frac{\dot{\phi}^2+\frac{3}{2}\frac{\dot{F}^2}{F}}{\left(H+\frac{1}{2}\frac{\dot{F}}{F}\right)^2}
\tilde{\gamma}_{\varphi,ij}
\frac{\partial_i HW_{\varphi}}{a\dot{\phi}}\frac{\partial_jHW_{\varphi}}{a\dot{\phi}}
\nonumber\\
&
+F\frac{\partial_i\partial_j\chi}{a^2}
\frac{\partial_k \tilde{\gamma}_{\varphi,ij}}{a}\frac{\partial_k \chi}{a}
+\frac{\partial_i\partial_j}{\nabla^2}\left[
\frac{\dot{\tilde{\gamma}}_{\varphi,ij}}{4}\frac{W_{\varphi}}{\dot{\phi}}
\frac{\frac{1}{2}\frac{\dot{F}}{F}}{H+\frac{1}{2}\frac{\dot{F}}{F}}\right]
\frac{2}{a^3}\frac{\delta S^{(2)}}{\delta W_{\varphi}}\left(-\frac{\dot{\phi}}{H}\right)
\nonumber\\
&
+\frac{\frac{1}{2}\frac{\dot{F}}{F}}{H+\frac{1}{2}\frac{\dot{F}}{F}}
\left[\frac{\frac{1}{2}\frac{\dot{F}}{F}}{H+\frac{1}{2}\frac{\dot{F}}{F}}
\frac{\partial_i W_{\varphi}}{a\dot{\phi}}\frac{\partial_j W_{\varphi}}{a\dot{\phi}}
-\left(\frac{\partial_i W_{\varphi}}{a\dot{\phi}}\frac{\partial_j \chi}{a}
+\frac{\partial_j W_{\varphi}}{a\dot{\phi}}\frac{\partial_i \chi}{a}\right)
\right]\frac{2}{a^3}\frac{\delta S^{(2)}}{\delta \tilde{\gamma}_{\varphi,ij}}
\Biggr\}
\,.
\label{nmaction:3:ssg:Wvarphi:2}
\end{align}
Similarly, the scalar-graviton-graviton actions
\eqref{nmaction:3:sgg:GI:Wzeta} and
\eqref{nmaction:3:sgg:GI:Wvarphi} become after
some partial integrations,
\begin{align}
S[W_{\zeta}\tilde{\gamma}_{\zeta}^2]=\frac12\int d^{3}xdt \bar{N} a^{3}
\Biggl\{&
\frac{\dot{\phi}^2+\frac{3}{2}\frac{\dot{F}^2}{F}}
{\left(H+\frac{1}{2}\frac{\dot{F}}{F}\right)^2}\frac{W_{\zeta}}{2}
\left(\frac14 \dot{\tilde{\gamma}}_{\zeta,ij}\dot{\tilde{\gamma}}_{\zeta,ij}
+\frac14\frac{\partial\tilde{\gamma}_{\zeta,ij}}{a}\frac{\partial\tilde{\gamma}_{\zeta,ij}}{a}\right)
\nonumber\\
&-\frac{F}{2} \dot{\tilde{\gamma}}_{\zeta,ij}\frac{\partial_k\tilde{\gamma}_{\zeta,ij}}{a}
\frac{\partial_k\chi}{a}
- \dot{\tilde{\gamma}}_{\zeta,ij}\frac{W_{\zeta}}{H+\frac{1}{2}\frac{\dot{F}}{F}}
\frac{2}{a^3}\frac{\delta S^{(2)}}{\delta \tilde{\gamma}_{\zeta,ij}}
\Biggr\}
\,,
\label{nmaction:3:sgg:GI:Wzeta:2}
\end{align}
and
\begin{align}
S[W_{\varphi}\tilde{\gamma}_{\varphi}^2]=\frac12\int d^{3}xdt \bar{N} a^{3}
\Biggl\{&
\frac{\dot{\phi}^2+\frac{3}{2}\frac{\dot{F}^2}{F}}
{\left(H+\frac{1}{2}\frac{\dot{F}}{F}\right)^2}\frac{1}{2}\left(\frac{-HW_{\varphi}}{\dot{\phi}}\right)
\left(\frac14 \dot{\tilde{\gamma}}_{\varphi,ij}\dot{\tilde{\gamma}}_{\varphi,ij}
+\frac14\frac{\partial\tilde{\gamma}_{\varphi,ij}}{a}\frac{\partial\tilde{\gamma}_{\varphi,ij}}{a}\right)
\nonumber\\
&-\frac{F}{2} \dot{\tilde{\gamma}}_{\varphi,ij}\frac{\partial_k\tilde{\gamma}_{\varphi,ij}}{a}
\frac{\partial_k\chi}{a}
-\dot{\tilde{\gamma}}_{\varphi,ij}\frac{-\frac{1}{2}\frac{\dot{F}}{F}}
{H+\frac{1}{2}\frac{\dot{F}}{F}}\frac{1}{H}\left(\frac{-HW_{\varphi}}{\dot{\phi}}\right)
\frac{2}{a^3}\frac{\delta S^{(2)}}{\delta \tilde{\gamma}_{\varphi,ij}}
\Biggr\}
\,.
\label{nmaction:3:sgg:GI:Wvarphi:2}
\end{align}
We see that the actions for $W_{\zeta}$ and $\tilde{\gamma}_{\zeta,ij}$
on the one hand, are almost of the same form of those for $W_{\varphi}$
and $\tilde{\gamma}_{\varphi,ij}$ on the other hand. They only differ
by terms proportional to the equation of motion and boundary terms
(which can be found in appendix \ref{sec: app: boundary terms}).
In fact, we can see that the actions \eqref{nmaction:3:ssg:Wzeta:2}
and \eqref{nmaction:3:ssg:Wvarphi:2} become of the same form
once we identify
\begin{align}
\tilde{\gamma}_{\varphi,ij}&=\tilde{\gamma}_{\zeta,ij}
-\dot{\tilde{\gamma}}_{\zeta,ij}\frac{W_{\zeta}}{H}
-\frac{\partial_i W_{\zeta}}{aH}\frac{\partial_j W_{\zeta}}{aH}
-\left(\frac{\partial_i W_{\zeta}}{aH}\frac{\partial_j s_{W_{\zeta}}}{a}
+\frac{\partial_j W_{\zeta}}{aH}\frac{\partial_i s_{W_{\zeta}}}{a}
\right)
\nonumber\\
-\frac{HW_{\varphi}}{\dot{\phi}}&=W_{\zeta}+\frac{\partial_i\partial_j}{\nabla^2}
\frac{1}{4}\dot{\tilde{\gamma}}_{\zeta,ij}\frac{W_{\zeta}}{H}
+\mathcal{O}(W_{\zeta}^2)
\,,
\label{WzetagammazetaJordanframerelations:2}
\end{align}
The action for the curvature and graviton perturbation
on uniform field hypersurfaces is thus related
to the action for the field and graviton perturbation on
uniform curvature hypersurfaces via non-linear field redefinitions.
Of course, Eqs. \eqref{WzetagammazetaJordanframerelations:2}
are nothing more than non-linear relations
between gauge invariant variables on different hypersurfaces.
It can be checked straightforwardly that the gauge invariant
variables \eqref{Wzeta:GI2ndorder} and \eqref{Wvarphi:gammavarphi:GI2ndorder}
satisfy the relation \eqref{WzetagammazetaJordanframerelations:2}.
The exercise in this section shows that the non-linear
relation between different hypersurfaces also follows
from the action. Note that we did not consider the boundary
terms here. They are discussed in the next section.

Some final words on uniqueness of the scalar-graviton interactions.
We have seen that the actions on different hypersurfaces differ
by terms proportional to the linear equations of motion and boundary terms.
This means that the complete non-linear equations of motion for $W_{\zeta}$ and
$\tilde{\gamma}_{\zeta,ij}$ are actually equal to those for $-H W_{\varphi}/\dot{\phi}$
and $\tilde{\gamma}_{\varphi,ij}$, since the boundary terms do not contribute
and the terms proportional to the equation of motion vanish at linear order.
Thus, the evolution of the fields, and of $n$-point functions, is the same
whether one works on one hypersurface or another. In that sense, there is a unique
bulk action that describes the evolution of non-Gaussianity \cite{Prokopec:2012ug}.
Still, $n$-point functions in the interaction can most certainly receive contributions from
boundary terms in the action, which we shall discuss next.

\subsubsection{3-point functions and boundary terms}
\label{sec: Boundary terms}
Due to the non-linear
relation between different variables on different hypersurfaces,
the 3-point functions on different hypersurfaces are related via
disconnected pieces. For example, in the case of pure scalar interactions,
the disconnected pieces are proportional to the square of the scalar power spectrum,
see for example \cite{Maldacena:2002vr}.
In the case of 3-point functions for scalar-graviton interactions, also
the 3-point function on one hypersurface differs from that on another
hypersurface by a product of the scalar and gravitational power spectrum.
For example, if we consider only the local contributions
in the non-linear relations \eqref{WzetagammazetaJordanframerelations:2},
we find that the scalar-graviton-graviton 3-point function on different
hypersurfaces are related as
\begin{align}
\langle \frac{-H W_{\varphi}(x_{1})}{\dot{\phi}} \tilde{\gamma}_{\varphi}(x_2)\tilde{\gamma}_{\varphi}(x_3)\rangle
&= \langle W_{\zeta}(x_1) \tilde{\gamma}_{\zeta}(x_2)\tilde{\gamma}_{\zeta}(x_3)\rangle
\nonumber\\
&-\frac{1}{H}\left( \langle W_{\zeta}(x_1)W_{\zeta}(x_2)\rangle
\langle\dot{\tilde{\gamma}}_{\zeta}(x_2)\tilde{\gamma}_{\zeta}(x_3) \rangle
+ \langle W_{\zeta}(x_1)W_{\zeta}(x_3)\rangle
\langle\tilde{\gamma}_{\zeta}(x_2)\dot{\tilde{\gamma}}_{\zeta}(x_3) \rangle
\right)+\ldots
\,.
\label{3pointfunctionrelation:sgg}
\end{align}
Here we used that $\tilde{\gamma}_{\varphi,ij}=\tilde{\gamma}_{\zeta,ij}
-\dot{\tilde{\gamma}}_{\zeta,ij}W_{\zeta}/H+\ldots$, and the dots
denote non-local contributions. Furthermore, we have made use
of the Wick contraction and the fact that $\langle W_{\zeta} \tilde{\gamma}_{\zeta}\rangle=0$.
The non-local terms are difficult to deal with in coordinate space,
but can be treated consistently in momentum space.
Considering the relation \eqref{3pointfunctionrelation:sgg}, one can show
that in the so-called squeezed limit -- where one of the momenta is much
smaller than the other two -- the scalar-graviton-graviton reduces
to a product of the scalar and graviton 2-point correlators
multiplied by the spectral index for tensor
fluctuations, $n_{\rm T}$, see Ref. \cite{Maldacena:2002vr}.
For the case of the scalar-scalar-graviton correlator, it
can be seen from the non-linear relations \eqref{WzetagammazetaJordanframerelations:2}
that the correlator on different hypersurfaces differs by non-local
terms.

Eq. \eqref{WzetagammazetaJordanframerelations:2}
is just an example of two gauge invariant variables that are
non-linearly related, but in principle it can be used for any non-linearly
related variables. This is very useful when computing 3-point functions
for a certain variable from an action, since one can choose to work
with a non-linearly related variable for which
the action takes a form which is most suitable in specific situations.
For example, during the slow-roll inflationary expansion it is useful to
use the action for $W_{\varphi}$ and $\tilde{\gamma}_{\varphi,ij}$, since
the orders in slow-roll are separated and it is easy to see what terms
are dominant. This is what was used in Ref. \cite{Maldacena:2002vr}
to compute the bispectrum for a minimally coupled scalar field.\\
\linebreak
Now some words on the boundary terms.
We have derived Eq. \eqref{3pointfunctionrelation:sgg} based purely
on the non-linear relations\eqref{WzetagammazetaJordanframerelations:2}.
Eq. \eqref{3pointfunctionrelation:sgg}
should also follow from the action, but this is not obvious when we
consider the rewritten actions on the uniform field hypersurface \eqref{nmaction:3:ssg:Wzeta:2}
and \eqref{nmaction:3:sgg:GI:Wzeta:2} versus the actions on uniform
curvature hypersurfaces \eqref{nmaction:3:ssg:Wvarphi:2}
and \eqref{nmaction:3:sgg:GI:Wvarphi:2}. The bulk actions
are of the same form, and therefore lead to similar
3-point functions. Terms proportional to the equations
of motion do not contribute to bispectrum, as they vanish
when evaluated on the solutions of the first order
equation of motion. This suggests
that the disconnected contributions should follow from
the boundary terms that are obtained after partial integrations.
Indeed, in the so-called in-in formalism temporal boundary terms
containing time derivatives of fields can contribute
to the bispectrum. These boundary terms were discussed
for the pure scalar case in Refs. \cite{Burrage:2011hd}\cite{Arroja:2011yj}
\cite{Rigopoulos:2011eq}\cite{Prokopec:2012ug}. There, it was found
that the temporal boundary terms precisely
provide the dominant contributions to the disconnected pieces
of the bispectrum. Thus, in the slow-roll regime
 the 3-point function for $W_{\zeta}$ can be computed \textit{via} the
3-point function for $W_{\varphi}$.

So, in the pure scalar case the non-linear field redefinition
\eqref{WzetagammazetaJordanframerelations:2}
takes into account the dominant contributions coming from the boundary
terms in the action for $W_{\zeta}$.
However, in previous work \cite{Prokopec:2012ug} we have argued
that the field redefinition should completely take care of all boundary
terms (by which the action on uniform field hypersurfaces differs
from that on uniform curvature hypersurfaces). The argument is
straightforward: the gauge invariant action for $W_{\zeta}$ and
$\tilde{\gamma}_{\zeta,ij}$, and the action for $W_{\varphi}$ and
$\tilde{\gamma}_{\varphi,ij}$, are both manifestly gauge invariant
actions at third order that follow from the same original covariant
action \eqref{nonminimalaction}. The variables are related via
a specific non-linear relation \eqref{WzetagammazetaJordanframerelations:2}.
Thus, also the manifestly gauge invariant actions should be related
by this non-linear relation. To be more precise, if we write the
non-linear transformations \eqref{WzetagammazetaJordanframerelations:2}
schematically as $\tilde{\gamma}_{\varphi,ij}=
\tilde{\gamma}_{\zeta,ij}+Q_{ij}(W_{\zeta},\tilde{\gamma}_{\zeta})$
and $-H W_{\varphi}/\dot{\phi}=
W_{\zeta}+Q_{ij}(W_{\zeta},\tilde{\gamma}_{\zeta})$
the quadratic actions
for $\tilde{\gamma}_{\varphi,ij}$ and $W_{\varphi}$
\eqref{nmaction:2:WvarphiWvarphi:GI}
change under the non-linear transformation as
\begin{align}
S^{(2)}[\tilde{\gamma}_{\varphi}^{2}]&=S^{(2)}[\tilde{\gamma}_{\zeta}^{2}]
+\frac12\int d^3xdt \bar{N} a^3\Biggl\{
Q_{ij}(W_{\zeta},\tilde{\gamma}_{\zeta})
\frac{2}{a^3}\frac{\delta S^{(2)}}{\delta \tilde{\gamma}_{\zeta,ij}}
\nonumber\\
&+\frac{1}{a^3}\left[
\frac{a^{3}F}{2}\dot{\tilde{\gamma}}_{\zeta,ij}
Q_{ij}(W_{\zeta},\tilde{\gamma}_{\zeta})
\right]^{\cdot}
-\frac{\partial_i}{a}\left[
\frac{F}{2}\frac{\partial_i\tilde{\gamma}_{\zeta,ij}}{a} Q_{ij}(W_{\zeta},\tilde{\gamma}_{\zeta})
\right]
\Biggr\}
\,,
\label{nmaction:2:gammagamma:nonlineartransform}
\end{align}
and
\begin{align}
S^{(2)}[W_{\varphi}^2]&=S^{(2)}[W_{\zeta}^2]
+\frac12\int d^3xdt \bar{N} a^3\Biggl\{
Q(W_{\zeta},\tilde{\gamma}_{\zeta})
\frac{2}{a^3}\frac{\delta S^{(2)}}{\delta W_{\zeta}}
\nonumber\\
&+\frac{1}{a^3}\left[
2a^{3}\frac{\dot{\phi}^2+\frac{3}{2}\frac{\dot{F}^2}{F}}
{\left(H+\frac{1}{2}\frac{\dot{F}}{F}\right)^2}\dot{W}_{\zeta}
Q(W_{\zeta},\tilde{\gamma}_{\zeta})
\right]^{\cdot}
-\frac{\partial_i}{a}\left[
2\frac{\dot{\phi}^2+\frac{3}{2}\frac{\dot{F}^2}{F}}
{\left(H+\frac{1}{2}\frac{\dot{F}}{F}\right)^2}
\frac{\partial_i W_{\zeta}}{a} Q(W_{\zeta},\tilde{\gamma}_{\zeta})
\right]
\Biggr\}
\,.
\label{nmaction:2:WvarphiWvarphi:nonlineartransform}
\end{align}
We have seen that the third order action for $W_{\zeta}$
and $\tilde{\gamma}_{\zeta,ij}$ can be written as the third
order action for $-H W_{\varphi}/\dot{\phi}$ and
$\tilde{\gamma}_{\varphi,ij}$, plus terms proportional to
the equation of motion and boundary terms, see
Eqs. \eqref{nmaction:3:ssg:Wzeta:2} and \eqref{nmaction:3:sgg:GI:Wzeta:2}.
The terms by which the actions differ are precisely
the equation of motion terms as expected by the
non-linear transformations
\eqref{nmaction:2:gammagamma:nonlineartransform}--\eqref{nmaction:2:WvarphiWvarphi:nonlineartransform}.
The boundary terms are however not precisely the expected ones.
In appendix \ref{sec: app: boundary terms} we give explicit expressions
for the boundary terms and show that the non-linear
the redefinitions \eqref{WzetagammazetaJordanframerelations:2}
do not completely take care of the boundary terms. Based on
gauge invariance of the action, all boundary terms must be taken
into account by the non-linear field redefinition.
We suspect that the discrepancy is due to two factors
that we did not consider. First, boundary terms
are already present in the second order action
\eqref{nmaction:2:wvarphiwvarphi:GI}. Under a non-linear
redefinition \eqref{WzetagammazetaJordanframerelations:2}
these second order boundary terms generate third order
boundary terms. Second, boundary terms are also
present in the ADM formulation
of the action, \textit{i.e.} by going from Eq. \eqref{nonminimalaction}
to Eq. \eqref{ADMactionnonminimalEij}. In total,
these boundary terms could make sure that the action on one hypersurface
is related to that on another, at the bulk as well
as at the boundary level.

\section{Frame independent cosmological perturbations}
\label{sec: Frame independent cosmological perturbations}
\subsection{The conformal transformation}
\label{sec: The conformal transformation}
So far we have worked exclusively in the Jordan frame and derived
the physical vertices for the manifestly gauge invariant variables.
It is however well known that the Jordan frame action with
non-minimal coupling \eqref{nonminimalaction}
can be brought to the Einstein frame action with minimal coupling
\begin{equation}
S=\frac12\int d^4 x \sqrt{-g_E}\left\{R_E
-g_E^{\mu\nu}\partial_{\mu}\Phi_E\partial_{\nu}\Phi_E -2V_E(\Phi_E)\right\}
\,.
\label{EF:action}
\end{equation}
by a combined redefinition of the metric and scalar field
\begin{align}
g_{\mu\nu,E}&=\Omega^2 g_{\mu\nu}
\nonumber\\
\left(\frac{d\Phi_E}{d\Phi}\right)^2
&=\frac{1}{\Omega^{2}}\left(1+6\Omega^{\prime 2}\right)
\nonumber\\
V_E(\Phi_E)&=\frac{1}{\Omega^4}V(\Phi)
\,,
\label{EinsteinJordanFrameRelations}
\end{align}
with
\begin{equation}
\Omega^2=\Omega^2(\Phi)=F(\Phi)
\,.
\end{equation}
No physical content is lost by field redefinitions,
and thus, in that sense, the Jordan and Einstein frame
are physically equivalent. Nonetheless, there is a
defining frame for \textit{e.g.} the Hubble parameter
that describes the rate of expansion of space.
Take the example of Higgs inflation \cite{Salopek:1988qh,Bezrukov:2007ep}.
Here the defining frame is the Jordan frame,
where $V(\Phi)$ is the Standard Model Higgs potential
and $\Phi$ is the Higgs field that is non-minimally
coupled to the Ricci scalar via $F=1+\xi\Phi^2$. The metric $g_{\mu\nu}$ describes
the curvature of spacetime on which the Higgs field lives.

Generally speaking, it is complicated to do computations directly in the Jordan
frame due to the non-minimal coupling.
Fortunately we can exploit the physical equivalence
of Jordan and Einstein frame. It is possible to first transform
to the Einstein frame, then perform the relevant
computations in that simpler frame, and finally transform
the results back to the Jordan frame.
For instance, for the background alone one can straightforwardly show
that
\begin{align}
\nonumber \bar{N}_E&=F^{\frac{1}{2}}\bar{N}\\
\nonumber a_E&=F^{\frac{1}{2}}a\\
\nonumber H_E&=\frac{1}{F^{\frac{1}{2}}}\left(H+\frac12\frac{\dot{F}}{F}\right)\\
\dot{\phi}_E&=\frac{1}{F^{\frac{1}{2}}}\frac{d\phi_E}{d\phi}\dot{\phi}
\equiv\frac{1}{F^{\frac{1}{2}}}
\sqrt{\frac{1}{F}+\frac{3}{2}\frac{F^{\prime 2}}{F^2}}\dot{\phi}
\,,
\label{EinsteinJordan Background Relations}
\end{align}
where the dotted derivative in Einstein frame is defined
as $\dot{\phi}_E=d\phi_E/(\bar{N}_E dt)$, and $H_E=\dot{a_E}/a_E$.
With these background relations the Friedman and field equations in
the Jordan frame \eqref{fieldequation} can be derived
immediately from the well-known Einstein frame equations,
\begin{align}
H_E^2&=\frac{1}{6}\left[\dot{\phi}_E^2+2V_E\right]
\nonumber\\
\dot{H}_E&=-\frac{1}{2}\dot{\phi}_E^2
\nonumber\\
0&=\ddot{\phi}_E+3H_E\dot{\phi}_E+V_E'
\label{fieldequation:Einstein}
\,.
\end{align}
This demonstrates the equivalence at the level
of the background equations of motion. In Higgs inflation,
or more general theories with non-minimal coupling,
the equivalence is frequently used to study the
inflationary expansion of the early universe. In
order to solve the horizon and flatness problems inflation
must have lasted sufficiently long, which translates
to the condition that the number of $e$-folds $N_e\equiv\int H \bar{N} dt$
exceeds 60. This means
that $H$ must be roughly constant and very slowly changing
in time, which is not easy to see in the Jordan frame. However,
in the Einstein frame it is easy to see that the potential
becomes almost flat for very large field values, which leads
to a slow-rolling trajectory and an approximately constant
Hubble parameter $H_E$, which in turn is closely related
to the Jordan frame Hubble parameter $H$
\footnote{It can be shown that in Higgs inflation the number
of $e$-folds in Jordan and Einstein frame are approximately
the same, $N_e\sim N_{e,E}$.}.

\subsection{Frame independent perturbations at first order}
\label{sec: Frame independent perturbations at first order}

Even though the equivalence between Jordan and Einstein frame
can be easily demonstrated for the background fields and action,
it is far from obvious for the perturbed action. The reason
is that the perturbations in Einstein frame do not coincide
with those in the Jordan frame. For example, if we expand
the complete field redefinitions \eqref{EinsteinJordanFrameRelations}
up to first order in perturbations, we find that
\begin{align}
\varphi_E&=\frac{d\phi_E}{d\phi}\varphi
\nonumber\\
\zeta_E&=\zeta+\frac12\frac{F'}{F}\varphi
\nonumber\\
\gamma_{ij,E}&=\gamma_{ij}
\,.
\label{JordanEinstein perturbation relation}
\end{align}
Thus, if we take the quadratic Einstein frame action
for perturbations $\varphi_E$, $\zeta_E$ and $\gamma_{ij,E}$,
with background quantities $H_E$ and $\dot{\phi}_E$, it is
difficult to see that we obtain the quadratic Jordan frame
action when we insert the relations \eqref{EinsteinJordan Background Relations}
and \eqref{JordanEinstein perturbation relation}.
Of course, we should obtain the perturbed Jordan frame action
after transforming the action in Einstein frame, since we
know that the complete Jordan and Einstein frame action
Eqs. \eqref{nonminimalaction} and \eqref{EF:action} are related
via (unperturbed) field redefinitions. Thus, the frame transformation
should also work order by order in perturbation theory.

At this point we can draw a nice analogy with the discussion
of gauge invariance in the previous section. For the original,
unperturbed action \eqref{nonminimalaction} general covariance
was manifest. Thus gauge invariance should also be present order
by order in perturbation theory. For the background the gauge
invariance is trivial, since the background is by definition
the part of the metric that does not transform under a gauge
transformation. The gauge invariance at the perturbative level
is not so obvious, because the perturbations themselves are
not gauge invariant. However, we were able to write the perturbed
action also in a manifestly gauge invariant way, by expressing
it in terms of variables which are inherently gauge invariant.

Based on this analogy we can ask ourselves: is it possible to express the
quadratic Einstein and Jordan frame
action in terms of certain variables in which
it is particularly simple to see that they are related
via a frame transformation? The answer is yes.
First of all, the graviton in the Einstein frame
is equal to that in the Jordan frame, see Eq.
\eqref{JordanEinstein perturbation relation}. Thus,
if we take the quadratic Einstein frame action
\begin{equation}
S^{(2)}[\gamma_E^2]=\frac12\int d^3xdt \bar{N}_E a_E^3\frac{1}{4}
\left[\dot{\gamma}_{ij,E}\dot{\gamma}_{ij,E}
-\left(\frac{\partial \gamma_{ij,E}}{a_E}\right)^2\right]
\,,
\label{EF:action:2:gammagamma}
\end{equation}
we immediately obtain the Jordan frame action \eqref{nmaction:2:gammagamma}
when we transform the background quantities according to Eq.
\eqref{EinsteinJordan Background Relations} and replace
$\gamma_{ij,E}=\gamma_{ij}$. Now, for the scalar perturbations
we can combine $\zeta_E$ and $\varphi_E$ in such a way that
\begin{equation}
w_{\zeta,E}=\zeta_E-\frac{H_E}{\dot{\phi}_E}\varphi_E
=\zeta-\frac{H}{\dot{\phi}}\varphi \equiv w_{\zeta}
\,.
\label{frame independence first order wzeta}
\end{equation}
Thus, the linearly gauge invariant curvature perturbation
in the Einstein frame coincides with the curvature
perturbation in the Jordan frame \cite{Makino:1991sg,Fakir:1992cg}.
Thus, in terms of this variable, the quadratic Jordan frame
action can be found straightforwardly from the Einstein frame
action by transforming the background quantities, as we
can see when we compare the Jordan frame action \eqref{nmaction:2:wzetawzeta:GI}
with the Einstein frame action
\begin{align}
S^{(2)}[w_{\zeta,E}^2]=\frac12\int d^{3}xdt \bar{N}_E a_E^{3}
\frac{\dot{\phi}_E^2}{H_E^2}
\left[\dot{w}_{\zeta,E}^2-\left(\frac{\partial w_{\zeta,E}}{a_E}\right)^2\right]
\,.
\label{EF:action:2:wzetawzeta:GI}
\end{align}
The gauge invariant graviton $\gamma_{ij}$ and curvature perturbation $w_{\zeta}$
take the same form after a frame transformation. Thus, they can be called
\textit{frame independent cosmological perturbations}, at least to first order.
Analogous to the discussion of gauge invariance, where
a manifestly gauge invariant form of the action is reached
when using gauge invariant variables, here
the action is "manifestly equivalent" in terms of frame independent
perturbations. What we mean is that, when expressed in terms of $w_{\zeta}$ and
$\gamma_{ij}$, the actions in Jordan and Einstein frame are directly
related via (trivial) frame transformations of the background alone.
Thus, if one computes certain results in the Einstein frame using
these frame independent perturbations, such as the scalar or tensorial
power spectrum, the corresponding Jordan frame result is immediately
obtained by transforming the background quantities. This was used in,
for example \cite{Salopek:1988qh,Makino:1991sg,Fakir:1992cg,Tsujikawa:2004my}.
Note that the power spectrum, for example, is expressed in terms of
$H$ and $\dot{\phi}$ in the Jordan frame, but in terms of $H_E$ and $\dot{\phi}_E$
in the Einstein frame. Of course, the actual \textit{value} of the amplitude and
spectral index is independent of the frame, once the equations of motion
for $H$ and $\dot{\phi}$, or $H_E$ and $\dot{\phi}_E$ are solved and their
values computed at horizon crossing. This shows that physical observables
are independent of the choice of frame.

It is no surprise that the gauge invariant curvature perturbation
on uniform field hypersurfaces is a frame independent perturbation
as well. In the gauge $\varphi=0$ the curvature perturbation $w_{\zeta}$
reduces to $\zeta$. Also, in this gauge the frame transformation, which
is a function of $\Phi$, becomes a function of the background
field alone. Thus the frame transformation does not affect
the perturbations, and the perturbed action in Jordan frame is simply
obtained from the Einstein frame by transforming the background fields alone.

\subsection{Frame independent perturbations at second order}

As we have seen in section ref{sec: Third order gauge invariant action},
at higher order the gauge transformations become non-linear, such
that the manifestly gauge invariant action at third order is expressed
in terms of second order gauge invariant variables. In analogy to this,
we expect that a manifestly equivalent
action can be expressed in terms of second order frame independent perturbations.
Based on the previous section, we are lead to believe that the second
order curvature perturbation and graviton on uniform field hypersurfaces
are also frame independent. This was verified previously for the scalar
perturbation $W_{\zeta}$ \cite{Sugiyama:2010zz,Prokopec:2012ug}, although
only the scalar gauge invariant part important on long wavelengths has been
considered. Here we computed the tensorial contributions to the second
order gauge invariant variable $W_{\zeta}$, and for those contributions
we can show using Eqs. \eqref{EinsteinJordan Background Relations} and
\eqref{JordanEinstein perturbation relation} that
\begin{align}
W_{\zeta,E}=w_{\zeta,E}
+\frac{\partial_i\partial_j}{\nabla^2}
\left(\frac{1}{4}\frac{\varphi_E}{\dot{\phi}_E}\dot{\gamma}_{ij,E}\right)
+\mathcal{O}(\varphi_E^2)
=w_{\zeta}
+\frac{\partial_i\partial_j}{\nabla^2}
\left(\frac{1}{4}\frac{\varphi}{\dot{\phi}}\dot{\gamma}_{ij}\right)
+\mathcal{O}(\varphi^2)
= W_{\zeta}
\label{Wzeta:secondorderframeindependent}
\,.
\end{align}
Similarly we can show for the graviton on uniform field hypersurfaces
\begin{align}
\tilde{\gamma}_{\zeta,ij,E}=&\gamma_{ij,E}
+\frac{\partial_i\varphi_E}{a_E\dot{\phi}_E}
\frac{\partial_j\varphi}{a_E\dot{\phi}_E}
-\left(\frac{\partial_i\varphi_E}{a_E\dot{\phi}_E}\frac{\partial_j s_{w_{\zeta,E}}}{a_E}
+\frac{\partial_j\varphi_E}{a_E\dot{\phi}_E}\frac{\partial_i s_{w_{\zeta,E}}}{a_E}
\right)
-\frac{\varphi_E}{\dot{\phi}_E}\dot{\gamma}_{ij}
\nonumber\\
=&\gamma_{ij}
+\frac{\partial_i\varphi}{a\dot{\phi}}
\frac{\partial_j\varphi}{a\dot{\phi}}
-\left(\frac{\partial_i\varphi}{a\dot{\phi}}\frac{\partial_j s_{w_{\zeta}}}{a}
+\frac{\partial_j\varphi}{a\dot{\phi}}\frac{\partial_i s_{w_{\zeta}}}{a}
\right)
-\frac{\varphi}{\dot{\phi}}\dot{\gamma}_{ij}
=\tilde{\gamma}_{\zeta,ij}
\,.
\label{gammaijzeta:secondorderframeindependent}
\end{align}
Here we used that $a_E^{-1}s_{w_{\zeta,E}}=a^{-1}s_{w_{\zeta}}$, where
$s_{w_{\zeta}}$ is defined in \eqref{lapsesolution:splitwzeta:nonminimal}.
Thus $W_{\zeta,E}=W_{\zeta}$ and $\tilde{\gamma}_{\zeta,ij,E}=\tilde{\gamma}_{\zeta,ij}$,
such that the gauge invariant scalar and tensor perturbations
on uniform field hypersurface are frame independent to second order
\footnote{The frame independence of the gauge invariant curvature perturbation can
be demonstrated to all orders \cite{Koh:2005ne,Chiba:2008ia} by
using the fully non-linear generalization of the curvature perturbation
\cite{Rigopoulos:2004gr,Langlois:2005ii,Lyth:2004gb}.}.
Of course, in the gauge $\varphi=0$ it becomes particularly clear
that $W_{\zeta}$ and $\tilde{\gamma}_{\zeta,ij}$ do not transform under a frame transformation,
which can be generalized to arbitrary order to show that the curvature perturbation
is frame independent \cite{Gong:2011qe}. By similar reasoning, also
the graviton on uniform field hypersurfaces is frame independent to all orders.

Now that we have proven that $W_{\zeta}$ and $\tilde{\gamma}_{\zeta,ij}$ are
frame independent variables, it is trivial to show that the cubic Jordan
and Einstein frame actions, expressed in those variables, are physically equivalent.
Take the cubic vertices for scalar-graviton interactions  in the Jordan frame,
Eqs.~\eqref{nmaction:3:ssg:GI:Wzeta} and \eqref{nmaction:3:sgg:GI:Wzeta}.
The same vertices are obtained by taking the corresponding Einstein frame
action (setting $F=1$) and transforming the background quantities
according to Eq. \eqref{EinsteinJordan Background Relations}. Thus one
can use the Einstein frame action to compute, for example, the 3-point function
for $W_{\zeta}$ in terms of $H_E$ and $\dot{\phi}$, and
then re-express everything in terms of $H$ and $\dot{\phi}$ to find
the 3-point function in the Jordan frame. Again, once these background
quantities are solved for through their equations of motion and their
values inserted in the $3$-point function, we would find exactly the same
number. For example, $f_{\rm NL}$ computed in the Einstein frame coincides
with that computed in Jordan frame, and is thus invariant under (unphysical)
field redefinitions \cite{Sugiyama:2010zz}.

Although the cubic action for $W_{\zeta}$ and $\tilde{\gamma}_{\zeta,ij}$
is "manifestly equivalent", the perturbative
physical equivalence of Jordan and Einstein
frame is not so obvious when the action is expressed in terms of the
gauge invariant field and graviton perturbations on uniform curvature
hypersurfaces, $W_{\varphi}$ and $\tilde{\gamma}_{\varphi,ij}$. The cubic interaction
vertices for these perturbations in the Einstein frame are
(set $F=1$ in Eqs.~\eqref{nmaction:3:ssg:GI:Wvarphi} and \eqref{nmaction:3:sgg:GI:Wvarphi})
\begin{align}
S[W_{\varphi,E}^2\tilde{\gamma}_{\varphi,E}]=\frac12\int d^{3}xdt \bar{N}_E a_E^{3}
&\Biggl\{
\frac{1}{2}\frac{\dot{\phi}_E^2}{H_E^2}
\frac{H_EW_{\varphi,E}}{\dot{\phi}_E}
\dot{\tilde{\gamma}}_{\varphi,ij,E}\frac{\partial_i\partial_j \chi_E}{a_E^2}
+\frac{\dot{\phi}_E^2}{H^2_E}\tilde{\gamma}_{\varphi,ij,E}
\frac{\partial_i H_EW_{\varphi,E}}{a_E\dot{\phi}_E}\frac{\partial_jH_EW_{\varphi,E}}{a_E\dot{\phi}_E}
\nonumber\\
&
+\frac{\partial_i\partial_j\chi_E}{a_E^2}
\frac{\partial_k \tilde{\gamma}_{\varphi,ij,E}}{a_E}\frac{\partial_k \chi_E}{a_E}
\Biggr\}
\,,
\label{EF:action:3:ssg:Wvarphi}
\end{align}
and
\begin{align}
S[W_{\varphi,E}\tilde{\gamma}_{\varphi,E}^2]=\frac12\int d^{3}xdt \bar{N}_E a_E^{3}
\Biggl\{&
\frac{\dot{\phi}_E^2}{H_E^2}\frac{1}{2}\left(\frac{-H_EW_{\varphi,E}}{\dot{\phi}_E}\right)
\left(\frac14 \dot{\tilde{\gamma}}_{\varphi,ij,E}\dot{\tilde{\gamma}}_{\varphi,ij,E}
+\frac14\frac{\partial\tilde{\gamma}_{\varphi,ij,E}}{a_E}\frac{\partial\tilde{\gamma}_{\varphi,ij,E}}{a_E}\right)
\nonumber\\
&-\frac{1}{2} \dot{\tilde{\gamma}}_{\varphi,ij,E}\frac{\partial_k\tilde{\gamma}_{\varphi,ij,E}}{a_E}
\frac{\partial_k\chi_E}{a_E}
\Biggr\}
\,.
\label{EF:action:3:sgg:Wvarphi}
\end{align}
Here $\chi_E$ is defined by
\begin{equation}
\frac{\nabla^2 \chi_E}{a_E^2}=\frac12 \frac{\dot{\phi}_E^2}{H_E^2}
\left(\frac{-H_EW_{\varphi,E}}{\dot{\phi}_E}\right)^{\cdot}
\,.
\end{equation}
Now, if we compare with the Jordan frame actions
~\eqref{nmaction:3:ssg:GI:Wvarphi} and \eqref{nmaction:3:sgg:GI:Wvarphi},
we see that the actions in the two frames are not simply
related by re-expressing the background fields as \eqref{EinsteinJordan Background Relations}.
The reason is that the second order field perturbation on uniform
curvature hypersurfaces, are not frame independent, $W_{\varphi,E}\neq W_{\varphi}$
and $\tilde{\gamma}_{\varphi,ij,E}\neq \tilde{\gamma}_{\varphi,ij}$. Even so,
it is not clear how exactly the different frames are related at the perturbative
level when working on the uniform curvature hypersurface.
Fortunately, we can make this more explicit by several partial integrations
of the Jordan frame action, which make it look more like
the transformed Einstein frame action. In fact, we have already performed
these partial integrations in section \ref{sec: Uniqueness of the scalar-graviton action},
and the resulting actions are Eqs. \eqref{nmaction:3:ssg:Wvarphi:2}
and \eqref{nmaction:3:sgg:GI:Wvarphi:2}. These actions are almost of
the same form as the Einstein frame actions \eqref{EF:action:3:ssg:Wvarphi}--\eqref{EF:action:3:sgg:Wvarphi}
after a frame transformation of the background,
but they differ by terms proportional
to the linear equation of motion and boundary terms.
In analogy to the relation between the actions on different hypersurfaces
in section \ref{sec: Uniqueness of the scalar-graviton action},
we can now identify
\begin{align}
\tilde{\gamma}_{\varphi,ij,E}=&\tilde{\gamma}_{\varphi,ij}
+\frac{\frac{1}{2}\frac{\dot{F}}{F}}{H+\frac{1}{2}\frac{\dot{F}}{F}}
\left[-\frac{W_{\varphi}}{\dot{\phi}}\dot{\tilde{\gamma}}_{\varphi,ij}
+\frac{\frac{1}{2}\frac{\dot{F}}{F}}{H+\frac{1}{2}\frac{\dot{F}}{F}}
\frac{\partial_iW_{\varphi}}{a\dot{\phi}}\frac{\partial_jW_{\varphi}}{a\dot{\phi}}
-\left(
\frac{\partial_iW_{\varphi}}{a\dot{\phi}}\frac{\partial_j\chi}{a}
+\frac{\partial_jW_{\varphi}}{a\dot{\phi}}\frac{\partial_i\chi}{a}\right)
\right]
\nonumber\\
\frac{H_EW_{\varphi,E}}{\dot{\phi}_E}
=&\frac{HW_{\varphi}}{\dot{\phi}}
-\frac{\frac{1}{2}\frac{\dot{F}}{F}}
{H+\frac{1}{2}\frac{\dot{F}}{F}}
\frac{\partial_i\partial_j}{\nabla^2}\left(\frac{1}{4}
\frac{W_{\varphi}}{\dot{\phi}}\dot{\tilde{\gamma}}_{\varphi,ij}\right)
+\mathcal{O}(W_{\varphi}^2)
\label{frametransformationssecondorder}
\,,
\end{align}
such that the Jordan and Einstein frame actions
on uniform curvature hypersurfaces are related by
combined transformations of the background \eqref{EinsteinJordan Background Relations}
and non-linear transformations of the perturbations \eqref{frametransformationssecondorder}
\footnote{The non-linear relations \eqref{frametransformationssecondorder}
can also be derived from the definitions of the gauge invariant perturbations
in the Einstein frame \eqref{Wvarphi:gammavarphi:GI2ndorder} and replacing
\eqref{EinsteinJordan Background Relations} and \eqref{JordanEinstein perturbation relation}.}.
This demonstrates the physical equivalence for the third order action
expressed in gauge invariant variables on the uniform curvature hypersurface,
although the action is not "manifestly equivalent".

The non-linear relation \eqref{frame independence first order wzeta}
between the gauge invariant variables in different frames
implies that the $n$-point functions in different frames
are related via disconnected terms. For example, the
3-point function for $W_{\varphi,E}$ differs from
the 3-point function for $W_{\varphi}$ by squares
of the 2-point function \cite{Prokopec:2012ug}. Thus, care must be taken when
computing Jordan frame quantities \textit{via} the Einstein
frame, if one works on the uniform curvature hypersurface.
An example of a delicate situation is the computation
of the naive cut-off in Higgs inflation
\cite{Lerner:2009na,Barbon:2009ya,Burgess:2010zq,Hertzberg:2010dc,Bezrukov:2010jz,Bezrukov:2011sz},
where the cut-off seems to depend on whether it is found directly
in the Jordan frame, or via the Einstein frame. As we have
discussed in previous work \cite{Prokopec:2012ug},
and emphasize again here, the cut-off should not
depend on the frame. This is most obvious when working
with frame independent variables, but the frame equivalence
is also non-linearly realized in the variables $W_{\varphi}$
and $\tilde{\gamma}_{\varphi,ij}$.

\section{Summary and outlook}

The aim of this work was to compute the gauge invariant
action at third order for a single scalar field in the Jordan frame,
with emphasis on the graviton and its interactions with the scalar
perturbations. Even though the unperturbed action is manifestly
covariant, it is not obviously so at the perturbative level.
The reason is that the perturbations themselves are gauge dependent.
We have demonstrated a method to find the manifestly gauge invariant
level order by order in perturbation theory. The procedure relies on
separating the higher order action in a gauge invariant plus
a gauge dependent part and absorbing the latter into the definition
of a new variable. By doing so one not only obtains the physical vertices,
but at the same time finds the correct higher order gauge invariant variables.
The method thus provides an alternative way to find gauge invariant variables
directly from the action, without having to resort to non-linear gauge
transformations.

In section \ref{sec: The gauge invariant action in the Jordan frame}
we computed the cubic gauge invariant action for the second
order curvature perturbation and graviton on uniform field hypersurfaces,
and for the second order field perturbation and graviton on
uniform curvature hypersurfaces, both directly in the Jordan frame.
We demonstrated that the different
action are related via non-linear transformations of the gauge
invariant variables, which are precisely the non-linear relations
between the variables on different hypersurfaces. The actions
on different hypersurfaces only differ by terms proportional
to the equation of motion and boundary terms, such
that the evolution of an $n$-point function is the same
for one set of gauge invariant variables or another.
In this sense the gauge invariant action is unique. Still,
the $n$-point functions, for example the bispectrum, for
one set of variables differs from that for another
by disconnected pieces due to the non-linear relation.

The situation concerning gauge invariance is quite similar
to that of the frame transformation. For the unperturbed action
in the Jordan frame it is well known that it can be brought
to the Einstein frame action via field redefinitions, and are in that
sense physically equivalent. This is however
not obvious at the perturbative level, since the perturbations
themselves are not invariant under the frame transformation.
However, the perturbed action can be written in a "manifestly
equivalent" form by expressing it in terms of frame independent
perturbations. We have demonstrated that the gauge invariant
curvature perturbation and graviton on uniform field hypersurfaces
in the Jordan frame coincide with the corresponding perturbations
in the Einstein frame, and are thus frame independent. In terms
of these variables, the perturbed action in the Jordan frame
can be obtained from the Einstein frame action by a frame
transformation of the background fields alone.
Moreover, we have shown that the field perturbation and graviton
on uniform curvature hypersurfaces in the Einstein frame are
non-linearly related to their counterparts in the Jordan frame.
This can be derived from the pertubations themselves, but also
follows from the action for these variables. As a consequence,
$n$-point functions for perturbations on uniform curvature
hypersurfaces differ by disconnected pieces between different
frames.

In conclusion, we have shown that whether one takes
the action for gauge invariant perturbations
on different hypersurfaces in the same frame, or the
action for the same gauge invariant perturbations
in different frames, they are all related via
non-linear transformations, which makes it
very convenient to find $n$-point functions
on a specific hypersurface/in a specific frame
\textit{via} the $n$-point functions on a
different hypersurface/in a different frame.

The results in this work may be used to compute
gauge invariant 3-point functions for scalar-graviton
interactions, and can be extended to compute gauge invariant 1-loop
quantum corrections to, for example, the scalar
or tensorial power spectrum in a general non-minimally coupled
theory. For such loop corrections one should derive
the fourth order gauge invariant action, and the procedure
outlined in this paper can be readily extended to do
precisely that. Moreover, we can readdress quantum
corrections and the naturalness problem in Higgs inflation
in an unambiguous way by using physical, frame independent
perturbations. We intend to address these questions
in following work.

\section{Acknowledgements}
This research was supported by the Dutch Foundation for
'Fundamenteel Onderzoek der Materie' (FOM) under the program
"Theoretical particle physics in the era of the LHC", program number FP 104.

\appendix

\section{Conventions and expansions}
\label{sec: Conventions and expansions}
In the main text we are using the action \eqref{ADMactionnonminimalEij}
and perturb it to third order using the second order perturbations
defined in Eq. \eqref{perturbations} with the spatial gauge fix
\eqref{spatial gauge fix}. Thus the perturbed metric becomes
\begin{equation}
g_{ij}=a^2e^{2\zeta}\left(e^{\gamma}\right)_{ij}
=a^2e^{2\zeta}\left(\delta_{ij}+\gamma_{ij}+\frac12 \gamma_{ik}\gamma_{kj}\right)
\,,
\label{metricperturbed:tensor}
\end{equation}
where we have expanded to second order. The conventions we are using here
is that, at the perturbative level, we define all quantities such as
spatial derivatives, tensors, vectors and the Kronecker delta with lower indices.
Thus, the Laplacian for perturbations is defined as $\nabla^2=\delta_{ij}\partial_i\partial_j$,
and the inverse of the metric is
\begin{equation}
g^{ij}=a^{-2}e^{-2\zeta}\left(e^{-\gamma}\right)_{ij}
=a^{-2}e^{-2\zeta}\left(\delta_{ij}-\gamma_{ij}+\frac12 \gamma_{ik}\gamma_{kj}\right)
\,.
\label{metricinverseperturbed}
\end{equation}
It can be checked straightforwardly that $g_{ik}g^{kj}=\delta_{ij}$.
Moreover, we define the shift perturbation with an upper index according
to \eqref{perturbations}. For the unperturbed quantities such as $N^{i}$ and
$E_{ij}$ indices are raised and lowered by the unperturbed metric,
\textit{i.e.} $N_i=g_{ij}N^j$ and $E^{ij}=g^{ia}g^{jb}E_{ab}$.

With the definition of the perturbed metric \eqref{metricperturbed:tensor}
the determinant becomes
\begin{equation}
\sqrt{g}=a^3e^{3\zeta}
\,.
\label{determinantgij}
\end{equation}
Derivatives of the metric are also particularly simple,
and at second order they are
\begin{align}
\dot{g}_{ij}&= 2(H+\dot{\zeta})g_{ij}+\frac12 (\dot{\gamma}_{ik}g_{kj}+g_{ik}\dot{\gamma}_{kj})
\nonumber \\
\partial_k g_{ij}&= 2\partial_k\zeta g_{ij}
+\frac12 (g_{lj}\partial_k \gamma_{il}+g_{il}\partial_k \gamma_{lj})
\,.
\end{align}
With this we can construct the Christoffel symbols
\begin{align}
\Gamma_{ij}^k&=\left[\delta_{ik}\partial_j \zeta+\delta_{jk}\partial_i\zeta -\delta_{ij}\partial_k \zeta\right]
\nonumber\\
&+\frac14\left[\partial_j\gamma_{ik}+\partial_i\gamma_{jk}
+g^{km}g_{il}(\partial_j \gamma_{lm}-\partial_m \gamma_{lj})
+g^{km}g_{jl}(\partial_i \gamma_{lm}-\partial_m \gamma_{li})\right]
\nonumber\\
\Gamma_{ik}^k&=3\partial_i\zeta
\,,
\label{Christoffelsymbolfirstorder}
\end{align}
which can be used to find for example
\begin{align}
\nabla_iN^{i}&=\partial_i N^{i}+3\partial_i\zeta N^{i}
\,,
\label{covariantderivativeexpansions}
\end{align}
and the Ricci scalar
\begin{align}
R=&-4g^{ij}\partial_i\partial_j\zeta - 2 g^{ij} \partial_i\zeta\partial_j\zeta
+2g^{ij}\partial_i\gamma_{jk}\partial_k\zeta
-\frac18 g^{ij} \partial_i\gamma_{kl}\partial_j\gamma_{kl}
-\frac18 g^{ij}\partial_k\gamma_{il}\partial_l\gamma_{jk}
\nonumber\\
&+\frac18 g^{ij}g^{kl}g_{mn}(-\partial_i \gamma_{km}\partial_j \gamma_{ln}
+\partial_i \gamma_{ln}\partial_k\gamma_{jm})
\,.
\label{Ricciscalar2}
\end{align}
If we expand this to third order in perturbations we find
\begin{align}
\sqrt{g}R=&ae^{\zeta}\biggl[-4\nabla^2\zeta - 2(\partial\zeta)^2
+4\gamma_{ij}\partial_i\partial_j\zeta+2\gamma_{ij}\partial_i\zeta\partial_j\zeta
-2\gamma_{ik}\gamma_{kj}\partial_i\partial_j\zeta-2\gamma_{ik}\partial_{i}\gamma_{kj}\partial_j\zeta
\nonumber\\
&-\frac14 \partial_i\gamma_{kl}\partial_i\gamma_{kl}
+\frac14 \gamma_{ij}\partial_i\gamma_{kl}\partial_j\gamma_{kl}
+\frac14 \gamma_{kl}\partial_i\gamma_{kj}\partial_j\gamma_{il}
-\frac14 \gamma_{ik}\partial_i\gamma_{jl}\partial_j\gamma_{kl}
\biggr]
\,.
\label{Ricciscalar3}
\end{align}

\section{Decoupling the constraint fields in the action}
\label{sec: Decoupling the constraint fields in the action}
We now present an alternative method to deal with the auxiliary
fields in the action. Instead of solving for them to first order
in perturbations, as was done in \cite{Maldacena:2002vr},
we can decouple them from the dynamical fields in the action.
The method was explained in Refs. \cite{Prokopec:2010be,Weenink:2010rr}.
After inserting the perturbations \eqref{perturbations} in the
action \eqref{ADMactionnonminimalEij} and expanding to second
order, we can collect only those terms that contain
perturbations of the lapse and shift field. After performing some
partial integrations and making use of the background equations
of motion the action takes the form
\begin{equation}
S^{(2)}[n,s]=\frac12\int d^{3}xdt \bar{N} a^{3}
\Biggl\{-2V n^2 + I_n n
+ 2 \left(2 H F +\dot{F}\right) n_{\rm sol} \frac{\nabla^2 s}{a^2}
\Biggr\}
\,,
\label{app:nmaction:2:  shift and lapse}
\end{equation}
where
\begin{align}
I_n =& 6(2HF +\dot{F})\dot{\zeta}+(6HF'-\dot{\phi})\dot{\varphi}
+(6H^2F'+6HF''-2V')\varphi
\nonumber\\
&-2 (2HF +\dot{F})\frac{\nabla^2s}{a^2}-2\frac{\nabla^2}{a^2}(2F\zeta+F'\varphi)
\,,
\end{align}
and
\begin{align}
n_{\rm sol}=\frac{1}{2HF+\dot{F}}\left[2F\dot{\zeta}+\dot{\phi}\varphi
+F'\dot{\varphi}-HF'\varphi+F''\dot{\phi}\varphi\right]
\,.
\end{align}
In terms of $n_{\rm sol}$ we may re-express $I_n$ as
\begin{equation}
I_n= 4 V n_{\rm sol} -2 (2HF+\dot{F}) \left(\frac{\nabla^2s}{a^2}-\frac{\nabla^2s_{\rm sol}}{a^2}\right)
\,,
\end{equation}
where
\begin{align}
\frac{\nabla^2s_{\rm sol}}{a^2}=&
\frac{-1}{H+\frac12\frac{\dot{F}}{F}}
\frac{\nabla^2}{a^2}
\left(\zeta+\frac12\frac{F'}{F}\varphi\right)
+\frac{1}{2F}\frac{\dot{\phi}^2+\frac{3}{2}\frac{\dot{F}^2}{F}}
{\left(H+\frac12\frac{\dot{F}}{F}\right)^2}
\left(\zeta-\frac{H}{\dot{\phi}}\varphi\right)^{\cdot}
\,.
\end{align}
Now we define new variables $\tilde{s}$ and $\tilde{n}$ as
\begin{align}
\frac{\nabla^2\tilde{s}}{a^2}&\equiv\frac{\nabla^2s}{a^2}-\frac{\nabla^2s_{\rm sol}}{a^2}
\nonumber\\
\tilde{n}&\equiv n - \frac{1}{4V} I_n = n- n_{\rm sol} -2 (2HF+\dot{F})\frac{\nabla^2\tilde{s}}{a^2}
\,.
\label{app:firstorderGIlapseshift}
\end{align}
In terms of these variables, the action \eqref{app:nmaction:2:  shift and lapse}
can be rewritten as
\begin{equation}
S^{(2)}[n,s]=\frac12\int d^{3}xdt \bar{N} a^{3}
\Biggl\{-2V \tilde{n}^2 + \frac{(2HF+\dot{F})^2}{V}\left(\frac{\nabla^2\tilde{s}}{a^2}\right)^2
+2V n_{\rm sol}^2 + 2 \left(2 H F +\dot{F}\right) n_{\rm sol} \frac{\nabla^2 s_{\rm sol}}{a^2}
\Biggr\}
\,.
\label{app:nmaction:2:  shift and lapse:2}
\end{equation}
Now, the first two terms in the action are the completely
decoupled quadratic actions for $\tilde{n}$ and $\tilde{s}$.
The equations of motions for these variables are simply
$\tilde{n}=0$ and $\nabla^2 \tilde{s}/a^2 =0$, which give
the first order solutions of the constraint equations,
$n=n_{\rm sol}$ and $\nabla^2s/a^2=\nabla^2 s_{\rm sol}/a^2$.
The additional two terms in the rewritten quadratic action are
precisely the terms that we would have obtained if we would
have solved the constraint equations to first order in perturbations
and inserted the solution in the action. This is what we wanted
to proof, although we have so far only shown it for the second
order action.

The extra terms in Eq. \eqref{app:nmaction:2:  shift and lapse:2}
are needed to construct the manifestly gauge invariant action for $w_{\zeta}$.
This implies that the variables $\tilde{n}$ and $\tilde{s}$ are
gauge invariant by themselves, which can be shown explicitly
\cite{Weenink:2010rr}. The gauge invariance can be exploited to
decouple non-dynamical degrees of freedom from dynamical ones
at higher order. The reasoning is very similar to that in
section \ref{sec: Scalar-scalar-graviton vertices on uniform field hypersurfaces}.

Let us consider the schematic form of the cubic action for the lapse
perturbation alone
\begin{equation}
S^{(3)}[n]=\frac12 \int d^{3}xdt\bar{N}a^{3}
\Biggl\{ 2V n^3 + n^2 Q^{(1)}+ n Q^{(2)}
\Biggr\}
\,.
\label{app:lapsecubicaction}
\end{equation}
Here the $Q^{(1)}$ and $Q^{(2)}$ are linear and
quadratic functions of the dynamical perturbations $\zeta$,
$\varphi$ and $\gamma_{ij}$ (and of $s$, but for simplicity
we neglect it). The explicit form of these
functions can be derived when expanding Eq. \eqref{ADMactionnonminimalEij}
up to third order in perturbations. We want to decouple the
non-dynamical degrees of freedom from the dynamical ones,
but this naively does not seem possible. Here is where
we exploit the gauge invariance of the action. Although $\tilde{n}$
is gauge invariant to first order, it transforms under
second order gauge transformations as
\begin{equation}
\tilde{n}\rightarrow \tilde{n}+\Delta_2^{\xi}\tilde{n}
\,.
\end{equation}
This induces a change in the quadratic action
\eqref{app:nmaction:2:  shift and lapse:2}
\begin{equation}
S^{(2)}[\tilde{n}]\rightarrow S^{(2)}[\tilde{n}]
+\frac12 \int d^{3}xdt \bar{N}a^3
\Biggl\{
\frac{2}{a^3}\frac{\delta S^{(2)}}{\delta \tilde{n}}\Delta^{\xi}_2\tilde{n}
\Biggr\}
= S^{(2)}[\tilde{n}] +\frac12 \int d^{3}xdt \bar{N}a^3
\Biggl\{
-4V\tilde{n}\Delta^{\xi}_2\tilde{n}
\Biggr\}
\,.
\end{equation}
Such a second order gauge transformation was also shown
for the dynamical perturbations \eqref{gaugetransformationsactionwzetagamma},
and a similar gauge transformation can be derived for the action for $\tilde{s}$.
Now, the only way in which such a second order gauge
transformation of the quadratic action can be balanced, is by terms proportional
to the equations of motion for $\tilde{n}$ in the third order
action. Since the equation of motion for a non-dynamical
field is simply proportional to the field itself,
we should find all the terms in the third order action
proportional to $\tilde{n}$. We can do this systematically for
the action \eqref{app:lapsecubicaction} by replacing
$n$ by $\tilde{n}$ using Eq. \eqref{app:firstorderGIlapseshift}.
The result is
\begin{align}
S^{(3)}[\tilde{n}]=\frac12 \int d^{3}xdt \bar{N}a^3&
\Biggl\{2V\tilde{n}^3
-4 V \tilde{n}\biggl[-\frac32 \tilde{n}n_{\rm sol}-\frac32 n_{\rm sol}^2
-\frac{Q^{(1)}\tilde{n}}{4V}-\frac{Q^{(1)}n_{\rm sol}}{2V}
-\frac{Q^{(2)}}{4V}\biggr]
\nonumber \\
& +  2V n_{\rm sol}^3 + n_{\rm sol}^2 Q^{(1)}+ n_{\rm sol} Q^{(2)}
\Biggr\}
\,.
\label{app:cubicactionlapsetilden}
\end{align}
The terms on the first line is a gauge invariant vertex for the
non-dynamical $\tilde{n}$, and terms proportional to the linear
equation of motion. The latter can be absorbed into the
quadratic action by defining a new second order gauge invariant variable
\begin{equation}
\tilde{N}=\tilde{n}-\frac32 \tilde{n}n_{\rm sol}-\frac32 n_{\rm sol}^2
-\frac{Q^{(1)}\tilde{n}}{4V}-\frac{Q^{(1)}n_{\rm sol}}{2V}
-\frac{Q^{(2)}}{4V}
\,,
\label{app:GaugeInvariantLapse}
\end{equation}
such that the cubic action \eqref{app:cubicactionlapsetilden}
becomes

\begin{align}
S^{(3)}[\tilde{N}]=\frac12 \int d^{3}xdt \bar{N}a^3&
\Biggl\{2V\tilde{N}^3
+  2V n_{\rm sol}^3 + n_{\rm sol}^2 Q^{(1)}+ n_{\rm sol} Q^{(2)}
\Biggr\}
\,.
\label{app:cubicactionlapsetilden}
\end{align}
Thus we have found a decoupled cubic vertex for $\tilde{N}$,
plus remaining terms which are the same as those
obtained by replacing $n$ in Eq. \eqref{app:lapsecubicaction}
by its first order solution.
In Eq. \eqref{app:cubicactionlapsetilden} we have neglected
the $\tilde{s}$ terms, but in a very similar way they can be
absorbed into the definition of a second order gauge
invariant shift perturbation $\tilde{S}$. So, finally
we are left with a decoupled, manifestly gauge invariant
cubic action for the non-dynamical fields $\tilde{N}$ and $\tilde{S}$,
and extra terms that we would have obtained when
the constraints would have been replaced by their first order
solution. This is precisely what we wanted to prove.
The method outlined here is a nice and systematic way to
decouple the non-dynamical sector from the dynamical one,
while at the same time finding the second order gauge
invariant constraint fields (which are the second
order solutions of the constraint equations).

The procedure here can be extended to higher order,
for example fourth order.
First, we found the second order gauge invariant
action in terms of the linearly gauge invariant
$\tilde{n}$ and $\tilde{s}$. Next, we insert these
quantities in the third and fourth order action. From
the third order action we now find the definition for
second order perturbations $\tilde{N}$ and $\tilde{S}$.
We can now replace in the third and fourth order action
$\tilde{n}=\tilde{N}+n_{\rm sol}^{(2)}+\ldots$
and $\tilde{s}=\tilde{S}+s_{\rm sol}^{(2)}+\ldots$,
where $n_{\rm sol}^{(2)}$ and $s_{\rm sol}^{(2)}$
are the second order parts of the solution of the constraint equation.
This gives gauge invariant cubic vertices for $\tilde{N}$
and $\tilde{S}$. Finally, the terms proportional to $\tilde{N}$ and $\tilde{S}$
in the fourth order action
can be absorbed in the quadratic action for $\tilde{N}$
and $\tilde{S}$ by defining
a third order gauge invariant variable. The remaining
terms in the action are those where $n$ and $s$ are replaced
by their second order solution $n_{\rm sol}^{(2)}$
and $s_{\rm sol}^{(2)}$. This was
also found in Ref. \cite{Chen:2006nt}, and applied in
Refs. \cite{Sloth:2006az,Seery:2006vu} to find the fourth
order action for perturbations in the uniform curvature gauge.

\section{Boundary terms}
\label{sec: app: boundary terms}

Here we discuss some of the boundary terms that appear due to the
partial integrations section \ref{sec: Uniqueness of the scalar-graviton action}.
We only show the temporal boundary terms, as these are
terms that can possibly contribute to the bispectrum.

We start with the boundary terms in the scalar-graviton-graviton
action. By going from Eq. \eqref{nmaction:3:sgg:GI:Wzeta}
to Eq. \eqref{nmaction:3:sgg:GI:Wzeta:2}, we find
the following temporal boundary terms
\begin{align}
S_{\partial}[W_{\zeta}\tilde{\gamma}_{\zeta}^2]=\frac12\int d^{3}x
\Biggl\{
\frac{a^{3}F}{H+\frac{1}{2}\frac{\dot{F}}{F}}W_{\zeta}
\left(-\frac14 \dot{\tilde{\gamma}}_{\zeta,ij}\dot{\tilde{\gamma}}_{\zeta,ij}
-\frac14\frac{\partial\tilde{\gamma}_{\zeta,ij}}{a}\frac{\partial\tilde{\gamma}_{\zeta,ij}}{a}\right)
\Biggr\}
\,.
\label{nmaction:3:sgg:GI:Wzeta:boundary}
\end{align}
Likewise, going from Eq. \eqref{nmaction:3:sgg:GI:Wvarphi}
to Eq. \eqref{nmaction:3:sgg:GI:Wvarphi:2} we obtain
\begin{align}
S_{\partial}[W_{\varphi}\tilde{\gamma}_{\varphi}^2]=\frac12\int d^{3}x
\Biggl\{&
\frac{a^{3}F}{H+\frac{1}{2}\frac{\dot{F}}{F}}
\frac{-\dot{F}}{2F}\frac{1}{H}\left(\frac{-HW_{\varphi}}{\dot{\phi}}\right)
\left(-\frac14 \dot{\tilde{\gamma}}_{\varphi,ij}\dot{\tilde{\gamma}}_{\varphi,ij}
-\frac14\frac{\partial\tilde{\gamma}_{\varphi,ij}}{a}\frac{\partial\tilde{\gamma}_{\varphi,ij}}{a}\right)
\Biggr\}
\,.
\label{nmaction:3:sgg:GI:Wvarphi:boundary}
\end{align}
In section \ref{sec: Boundary terms} it was argued that
by the non-linear relation \eqref{WzetagammazetaJordanframerelations:2}
the action for $W_{\varphi}$ and $\tilde{\gamma}_{\varphi}$
should transform into that for $W_{\zeta}$ and $\tilde{\gamma}_{\zeta}$,
both at the bulk and boundary level. If we consider
the scalar-graviton-graviton action alone, and see
what boundary terms for $W_{\zeta}$ and $\tilde{\gamma}_{\zeta}$
we get under the transformation
\eqref{WzetagammazetaJordanframerelations:2}, we find
\begin{align}
\frac12\int d^{3}x
\Biggl\{&
\frac{a^{3}F}{H+\frac{1}{2}\frac{\dot{F}}{F}}
\frac{-\dot{F}}{2F}\frac{W_{\zeta}}{H}
\left(-\frac14 \dot{\tilde{\gamma}}_{\zeta,ij}\dot{\tilde{\gamma}}_{\zeta,ij}
-\frac14\frac{\partial\tilde{\gamma}_{\zeta,ij}}{a}\frac{\partial\tilde{\gamma}_{\zeta,ij}}{a}\right)
-\frac{a^3F}{2}\dot{\tilde{\gamma}}_{\zeta,ij}\frac{W_{\zeta}}{H}\dot{\tilde{\gamma}}_{\zeta,ij}
\Biggr\}
\nonumber\\
&=S_{\partial}[W_{\zeta}\tilde{\gamma}_{\zeta}^2]
+\frac12\int d^{3}x
\Biggl\{
a^{3}F\frac{W_{\zeta}}{H}
\left(-\frac14 \dot{\tilde{\gamma}}_{\zeta,ij}\dot{\tilde{\gamma}}_{\zeta,ij}
+\frac14\frac{\partial\tilde{\gamma}_{\zeta,ij}}{a}\frac{\partial\tilde{\gamma}_{\zeta,ij}}{a}\right)
\Biggr\}
\,.
\label{nmaction:3:sgg:diffboundary}
\end{align}
The first line contains terms coming from the boundary terms
in Eq. \eqref{nmaction:3:sgg:GI:Wvarphi:boundary}, plus
terms that are generated from the quadratic action through
the transformation, see Eq. \eqref{nmaction:2:gammagamma:nonlineartransform}.
Eq. \eqref{nmaction:3:sgg:diffboundary} suggests
that the scalar-graviton-graviton boundary terms
are not related via the non-linear transformation
\eqref{WzetagammazetaJordanframerelations:2}. In the
main text we argue why the boundary terms must be
related as well, and we discuss what may cause the discrepancy.\\

Similarly, we can compute the boundary terms
for scalar-scalar-graviton interactions.
By going from Eq. \eqref{nmaction:3:ssg:GI:Wzeta}
to Eq. \eqref{nmaction:3:ssg:Wzeta:2}, we find
the following temporal boundary terms
\begin{align}
S_{\partial}[W_{\zeta}^2\tilde{\gamma}_{\zeta}]=\frac12\int d^{3}x
\Biggl\{
\frac{-a^{3}F}{H+\frac12\frac{\dot{F}}{F}}
&\Biggl[
\dot{\tilde{\gamma}}_{\zeta,ij}\frac12
\left(\frac{\partial_iW_{\zeta}}{a}\frac{\partial_j\chi}{a}
+\frac{\partial_jW_{\zeta}}{a}\frac{\partial_i\chi}{a}
-\frac{1}{H+\frac12\frac{\dot{F}}{F}}
\frac{\partial_iW_{\zeta}}{a}\frac{\partial_jW_{\zeta}}{a}
\right)
\nonumber\\
&+4\tilde{\gamma}_{\zeta,ij}
\frac{\partial_i W_{\zeta}}{a}\frac{\partial_jW_{\zeta}}{a}
\Biggr]
\Biggr\}
\,.
\label{nmaction:3:ssg:Wzeta:boundary}
\end{align}
Likewise, after many partial integrations
to get Eq. \eqref{nmaction:3:ssg:Wvarphi:2}
from Eq. \eqref{nmaction:3:ssg:GI:Wvarphi},
we find the following boundary terms,
\begin{align}
S_{\partial}[W_{\varphi}^2\tilde{\gamma}_{\zeta}]=\frac12\int d^{3}x
\Biggl\{
\frac{a^{3}
\frac{1}{2}\frac{\dot{F}}{F}}{H+\frac{1}{2}\frac{\dot{F}}{F}}
&
\Biggl[
\frac{F}{2}\dot{\tilde{\gamma}}_{\varphi,ij}
\left(\frac{\frac{1}{2}\frac{\dot{F}}{F}}{H+\frac{1}{2}\frac{\dot{F}}{F}}
\frac{\partial_i W_{\varphi}}{a\dot{\phi}}\frac{\partial_j W_{\varphi}}{a\dot{\phi}}
-\frac{\partial_i W_{\varphi}}{a\dot{\phi}}\frac{\partial_j \chi}{a\dot{\phi}}
-\frac{\partial_j W_{\varphi}}{a\dot{\phi}}\frac{\partial_i \chi}{a\dot{\phi}}
\right)
\nonumber\\
&
-\dot{F}\tilde{\gamma}_{\varphi,ij}
\frac{\partial_i W_{\varphi}}{a\dot{\phi}}\frac{\partial_jW_{\varphi}}{a\dot{\phi}}
\Biggr]
\Biggr\}
\,.
\label{nmaction:3:ssg:Wvarphi:boundary}
\end{align}
Also here we find that, after the redefinition
of $W_{\varphi}$ and $\tilde{\gamma}_{\varphi}$
using Eq. \eqref{WzetagammazetaJordanframerelations:2}, the
total boundary terms for scalar-scalar-graviton
interactions do not agree with those in \eqref{nmaction:3:ssg:Wzeta:boundary}.

\bibliography{Higgsinflation}{}
\bibliographystyle{JHEP}

\end{document}